\documentclass[twocolumn]{autart}

\usepackage[utf8]{inputenc}
\usepackage[T1]{fontenc}
\usepackage{amsmath,amssymb}
\usepackage{enumerate}
\usepackage{graphicx}
\usepackage{xcolor}
\usepackage{algorithm,algorithmicx,algpseudocode}

\newlength{\plotwidth}
\newlength{\plotwidthfull}

\newcommand{\seq}[3]{\left\{#1_{i}\right\}_{i=#2}^{#3}}
\newcommand{\seqk}[3]{\left\{#1_{i|k}\right\}_{i=#2}^{#3}}
\newcommand{\restr}[3]{\mathcal{R}\left(#1\nmidsep #2,#3\right)}
\newcommand{\img}[3]{\mathcal{I}(#1,#2|#3)}

\newcommand{\idxo}[2]{\ensuremath{[#1..#2)}}
\newcommand{\idxcl}[2]{\ensuremath{[#1..#2]}}
\newcommand{\ub}[1]{\overline{#1}}
\newcommand{\lb}[1]{\underline{#1}}
\newcommand{\classL}{\ensuremath{\mathcal{L}}}
\newcommand{\classKL}{\ensuremath{\mathcal{KL}}}
\newcommand{\classK}{\ensuremath{\mathcal{K}}}
\newcommand{\classKinfty}{\ensuremath{\mathcal{K}_\infty}}
\newcommand{\midsep}{\ensuremath{\,\middle|\,}}

\newcommand{\nmidsep}{\ensuremath{\,|\,}}

\renewcommand{\qed}{\relax\ifmmode\hfill\square\else\hfill$\square$\fi}

\begin{document}

\begin{frontmatter}
    \title{Heterogeneously parameterized tube model predictive control for LPV systems\thanksref{footnoteinfo}}
    
    \thanks[footnoteinfo]{This paper was not presented at any conference.\\
    This work was supported by the Impulse 1 program of Eindhoven University of Technology and ASML.
    This work has received funding from the European Research Council (ERC) under the European Unions Horizon 2020 research and innovation programme (grant agreement No 714663).\\
    This work is licensed under the Creative Commons Attribution-NonCommercial-NoDerivs 2.0 Generic License. To view a copy of this license, visit http://creativecommons.org/licenses/by-nc-nd/2.0/ or send a letter to Creative Commons, PO Box 1866, Mountain View, CA 94042, USA.\\
        * Corresponding author J. Hanema.}
    
    \author[csgroup]{J. Hanema}\ead{jurre@jhanema.nl}$^{,\ast}$,
    \author[csgroup]{M. Lazar}\ead{m.lazar@tue.nl},
    \author[csgroup]{R. Tóth}\ead{r.toth@tue.nl}
    
    \address[csgroup]{Department of Electrical Engineering, Eindhoven University of Technology, Eindhoven, The Netherlands.}                         
    
    
    \begin{abstract}
        This paper presents a heterogeneously parameterized tube-based model predictive control (MPC) design applicable to linear parameter-varying (LPV) systems.
        In a heterogeneous tube, the parameterizations of the tube cross sections and the associated control laws are allowed to vary along the prediction horizon.
        Two extreme cases that can be described in this framework are scenario MPC (high complexity, larger domain of attraction)
        and homothetic tube MPC with a simple time-invariant control parameterization (low complexity, smaller domain of attraction).
        In the proposed framework, these extreme parameterizations, as well as other parameterizations of intermediate complexity, can be combined within a single tube.
        By allowing for more flexibility in the parameterization design,
        one can influence the trade-off between computational cost and the size of the domain of attraction.
        Sufficient conditions on the parameterization structure are developed under which recursive feasibility and closed-loop stability are guaranteed.
        A specific parameterization that combines the principles of scenario and homothetic tube MPC is proposed and it is shown to satisfy the required conditions.
        The properties of the approach, including its capability of achieving improved complexity/performance trade-offs, are demonstrated using two numerical examples.
    \end{abstract}
\end{frontmatter}


\section{Introduction}
\label{sec:introduction}

This paper considers model predictive control (MPC) of linear parameter-varying (LPV) systems that can be represented in the state-space form $x(k+1)=A(\theta(k))x(k)+B(\theta(k))u(k)$,
where $A(\cdot)$ and $B(\cdot)$ are affine matrix functions of $\theta$.
In an LPV system, the state transition map is linear, but this linear map depends on the external \emph{scheduling variable} denoted by $\theta$.
In this setting, the current value $\theta(k)$ can be measured for all times $k$, but the future behavior of $\theta$ is generally not known exactly at time $k$.
Solving a predictive control problem under uncertainty requires the on-line optimization over feedback policies, leading to a so-called min-max feedback control problem \cite{Lee1997}.
This problem can be solved using dynamic programming (DP) \cite{RawlingsMayne2009}, but typically this is computationally intractable in practice.

Therefore it is useful to search for more conservative, but implementable, approximations of this difficult problem \cite{Rakovic2015a}.
Frequently used approaches for the control of constrained LPV systems are based on the on-line synthesis of linear feedback policies, e.g.,
\cite{Lu2000,Casavola2002,Casavola2007,Casavola2008a,Zheng2013}.
Robust MPC for parametrically uncertain\footnote{In this paper, a system with the same mathematical structure as an LPV system, but with a non-measurable scheduling variable, is called a parametrically or multiplicatively uncertain system.} systems furthermore can be based, e.g., on interpolation \cite{Bacic2003,Pluymers2005f,Wan2006} or on lifted ``prediction dynamics''
\cite{Kouvaritakis2000,Cannon2005}.
In this paper, the focus is on a different paradigm devised to reduce complexity with respect to the min-max solution, namely tube model predictive control (TMPC).
Compared to the approaches mentioned previously, an attractive feature of TMPC is that it allows for the use of arbitrary prediction horizons with a computational complexity that grows
linearly with the horizon length.
In the LPV case, the tube-based framework can be used to construct ``anticipative'' controllers, i.e., controllers that can take advantage of information on
possible future scheduling trajectories that becomes available while the system is running \cite{Hanema2016:cdc:final}.

Tube-based approaches were originally proposed to control constrained linear systems subject to \emph{additive} disturbances \cite{Langson2004,Mayne2005,Rakovic2012a,Rakovic2012,Rakovic2016a}.
The current paper however considers tube-based control of LPV systems,
where the uncertainty in the future evolution of the scheduling variable enters \emph{multiplicatively} instead of additively.
In \cite{Langson2004}, the authors discuss the possibility of adapting their TMPC to parametrically uncertain systems, but without investigating closed-loop stability.
Existing TMPC approaches for multiplicatively uncertain systems are, e.g., \cite{Munoz-Carpintero2013a,Fleming2015}.
A framework for the construction of ``stabilizing'' tubes, with application to the predictive control of linear systems on assigned initial condition sets, was presented in \cite{Brunner2013}.
An LPV TMPC based on the setting of \cite{Brunner2013} was presented in \cite{Hanema2016:cdc:final}:
therein, the constructed tubes are homothetic to the terminal set, and the on-line optimization of parameterized feedback policies is done over vertex controllers.
This approach was further extended in \cite{Hanema2017:aut:final}, which introduced relaxed finite-step terminal conditions into tube-based MPC.

The properties of a tube-based controller are determined to a large extent by the selected tube parameterization.
Specifically, the parameterization determines how well a tube-based controller can approximate the full DP solution.
A rich parameterization with many degrees of freedom (DOFs) makes it possible to achieve good control performance close to DP, but at a high associated computational cost.
On the other hand, a simple parameterization can lead to efficient optimization problems, but it limits the achievable performance.
Hence, a key question that motivates the work in this paper, is how to parameterize the tube to strike a good balance between computational complexity and control performance.

Typically, in the literature, a single tube parameterization is selected for the full prediction horizon:
e.g., homothetic tubes with vertex controls in \cite{Langson2004,Hanema2016:cdc:final} or elastic tubes with additional control actions superimposed onto a linear state feedback in \cite{Fleming2015}.
The restriction to one single parameterization for the full horizon limits the freedom that is available for the design of tube parameterizations that achieve favorable complexity/performance trade-offs.
In TMPC of linear time-invariant (LTI) systems subject to additive disturbances, this situation is mostly resolved because it is also possible to optimize over disturbance-feedback policies in a computationally efficient manner \cite{Goulart2008,Rakovic2012}.
However, in the case of an LPV model, the uncertainty enters multiplicatively,
and it is not possible to formulate the synthesis of disturbance-feedback policies as a convex optimization problem.

Therefore, to be able to construct tube-based controllers for LPV systems that can achieve better complexity/performance trade-offs,
new approaches for designing tube parameterizations are necessary.
To this end, as the first contribution of this paper, the concept of heterogeneously parameterized tubes (HpTs) is introduced.
In an HpT, the parameterization of the cross sections and associated controllers can vary along the prediction horizon.
This removes the restriction that one single parameterization must be selected for the full prediction horizon.
The new design freedom allowed by this framework can be exploited by the user to design tube parameterizations that achieve different improved complexity/performance trade-offs.
A number of parameterizations from the literature can be described in the proposed framework in a unified fashion,
and can be combined together to synthesize a single HpT.
Possible parameterizations that can be described in the HpT framework include homothetic- and elastic tubes \cite{Langson2004,Rakovic2012a,Rakovic2016a,Fleming2015},
but also so-called scenario tubes \cite{Lucia2013,Maiworm2015,Scokaert1998,Kerrigan2004,MunozDeLaPena2005,MunozDeLaPena2006}.
Based on the introduced HpT concept, a novel LPV MPC algorithm based on repetitive on-line construction of an HpT is developed.
It is worth to point out that the recent work \cite{Subramanian2018a} considers a different combination of scenario and tube-based MPC, i.e.,
by using a scenario tree to handle parametric uncertainties and a tube to handle additive disturbances.

The second and main contribution of the paper is the development of
sufficient conditions on the underlying heterogeneous parameterization, under which the resulting controller is recursively feasible and asymptotically stabilizing.
As the third contribution, an implementable heterogeneous parameterization---called HpT-SF---is proposed as a specific application of the developed general HpT framework.
This HpT-SF parameterization combines the principles of scenario and homothetic TMPC,
providing more design DOFs that can be leveraged to achieve improved complexity/performance trade-offs.

The remainder of this paper is structured as follows.
Section~\ref{sec:prelim} introduces the necessary preliminaries including notation, problem setting and the concept of scheduling tubes.
The concept of heterogeneously parameterized tubes (HpT) is presented in Section~\ref{sec:hpt}
and the TMPC algorithm based on these tubes is developed in Section~\ref{sec:tmpc}.
Conditions such that the algorithm is recursively feasible and stabilizing are given therein.
Subsequently, in Section~\ref{sec:design}, a terminal cost function and an implementable heterogeneous parameterization are provided that satisfy the required assumptions.
Numerical examples are provided in Section~\ref{sec:numex} to demonstrate that the HpT can potentially achieve improved complexity/performance trade-offs.
Concluding remarks are given in Section~\ref{sec:conclusion}.


\section{Preliminaries}
\label{sec:prelim}


\subsection{Notation and basic definitions}
\label{sec:prelim:notation}

The set of real numbers is denoted by $\mathbb{R}$ and the set of non-negative real numbers by $\mathbb{R}_+$.
Closed and open intervals on $\mathbb{R}$ are denoted by $[a,b]$ and $(a,b)$, respectively.
The symbol $\mathbb{N}$ is used to denote the set of non-negative integers (i.e., the integers including zero).
Closed and open index sets on $\mathbb{N}$ are defined as $\idxcl{a}{b} = \{i\in\mathbb{N}\,|\,a\leq i\leq b\}$ and
$\idxo{a}{b} = \{i\in\mathbb{N}\,|\,a\leq i< b\}$, respectively.
A set with a non-empty interior that contains the origin is called a proper set,
and a proper set which is also compact and convex is called a PC-set.
A polyhedron is a convex set that can be represented as the intersection of finitely many half-spaces.
A polytope is a compact polyhedron and can equivalently be described as the convex hull of finitely many vertices.
The power set of $\mathcal{A}\subseteq\mathbb{R}^n$ is the set of all subsets of $\mathcal{A}$ (including the empty set and $\mathcal{A}$ itself), and is denoted by $2^\mathcal{A}$.
Sequences are denoted compactly as $\seq{X}{a}{b} = \{X_a, X_{a+1}, \dots, X_b\}$.
The Minkowski sum of two sets $\mathcal{A}\subseteq\mathbb{R}^n$ and $\mathcal{B}\subseteq\mathbb{R}^n$ is
$\mathcal{A}\oplus \mathcal{B} = \left\{a+b\,|\,a\in \mathcal{A}, b\in \mathcal{B}\right\}$.
If $a\in\mathbb{R}^n$ is a vector, define $a\oplus\mathcal{B} = \left\{a+b\,|\,b\in \mathcal{B}\right\}$.
The $N$-times Cartesian product of a set $\mathcal{A}$ is $\mathcal{A}^N=\mathcal{A}\times\dots\times\mathcal{A}$.
The Hausdorff distance between two sets $\mathcal{A}\subseteq\mathbb{R}^n$ and $\mathcal{B}\subseteq\mathbb{R}^n$ is
\begin{equation*}
    d_H\left(\mathcal{A}, \mathcal{B}\right) = \max\bigl\{\sup_{a\in\mathcal{A}} \inf_{b\in\mathcal{B}} \|a-b\|, \sup_{b\in\mathcal{B}} \inf_{a\in\mathcal{A}} \|a-b\|\bigr\}
\end{equation*}
where $\|\cdot\|$ can be any vector norm on $\mathbb{R}^n$.
The Hausdorff distance between a set $\mathcal{A}\subseteq\mathbb{R}^n$ and the origin is therefore
\begin{equation}\label{eq:hausdorff-orig}
    d_H^0\left(\mathcal{A}\right) = d_H\left(\mathcal{A}, \{0\}\right) = \sup_{a\in\mathcal{A}} \|a\|.
\end{equation}
A function $f:\mathbb{R}_+\rightarrow\mathbb{R}_+$ is of class $\classK$ if it is continuous, strictly increasing, and $f(0)=0$.
It is in class $\classKinfty$ if, next to being in class $\classK$, $\lim_{\xi\rightarrow\infty}f(\xi)=\infty$.
A function $g:\mathbb{R}_+\rightarrow\mathbb{R}_+$ is of class $\classL$ if it is continuous, strictly decreasing, and $\lim_{\xi\rightarrow 0} g(\xi)=0$.
Lastly, a function $h:\mathbb{R}_+\times\mathbb{R}_+\rightarrow\mathbb{R}_+$ is said to be in class $\classKL$ if it is class-$\classK$ in its first argument and class-$\classL$ in its second argument.

Define the following ``set''-gauge function:

\begin{defn}\cite{Hanema2017:aut:final}\label{def:set-gauge}
    The set-gauge function
    $\Psi_S:2^{\mathbb{R}^n}\rightarrow \mathbb{R}_+$ corresponding to a PC-set $S\subset \mathbb{R}^n$ is
    \begin{equation*}
    \Psi_S(X) = \sup_{x\in X} \psi_S(x) = \inf\left\{\gamma\geq 0\mid X\subseteq \gamma S\right\}.
    \end{equation*}
\end{defn}



\subsection{Problem setting}
\label{sec:prelim:problem-setting}

We consider a constrained LPV system, represented by the following LPV state-space (LPV-SS) equation
\begin{equation}\label{eq:lpv-ss}
    x(k+1) = A(\theta(k))x(k) + B(\theta(k))u(k),\ k\in\mathbb{N},
\end{equation}
with the initial condition $x(0)=x_0$, and where $u:\mathbb{N}\rightarrow\mathbb{U}\subseteq\mathbb{R}^{n_\mathrm{u}}$ is the input, $x:\mathbb{N}\rightarrow\mathbb{X}\subseteq\mathbb{R}^{n_{\mathrm{x}}}$ is the state variable, and $\theta:\mathbb{N}\rightarrow\Theta\subseteq\mathbb{R}^{n_{\mathrm{\theta}}}$ is the scheduling signal.
The sets $\mathbb{U}$ and $\mathbb{X}$ are the input and state constraint sets, while $\Theta$ is called the scheduling set.
The matrices $A(\theta)$ and $B(\theta)$ in (\ref{eq:lpv-ss}) are affine functions of $\theta$, i.e.,
\begin{equation*}
    A(\theta) = A_0 + \sum_{i=1}^{n_\theta} \theta_i A_i,\quad
    B(\theta) = B_0 + \sum_{i=1}^{n_\theta} \theta_i B_i,
\end{equation*}
where $(A_i,B_i)$, $i\in\idxcl{0}{n_\theta}$, are conformable matrices.
The following standing assumptions are made.
\begin{assum}\label{ass:system-assumptions}
    The system represented by \eqref{eq:lpv-ss} satisfies:
    \begin{enumerate}[(i)]
        \item The values $x(k)$ and $\theta(k)$ can be measured at every time $k\in\mathbb{N}$.

        \item The sets $\mathbb{X}$ and $\mathbb{U}$ are polytopic PC-sets.
    \end{enumerate}
\end{assum}

The problem addressed in this paper is to design a controller $K_\mathrm{mpc}:\mathbb{X}\times\Theta\times\mathbb{N}\rightarrow\mathbb{U}$,
such that the origin is a regionally asymptotically stable equilibrium of the closed-loop system represented by
\begin{equation}\label{eq:lpv-cl}
    \begin{aligned}
        x(k + 1) &= A\left(\theta(k)\right)x(k) + B\left(\theta(k)\right)K_\mathrm{mpc}\left(x(k),\theta(k),k\right)\\
        &= \Phi\left(x(k), \theta(k), k\right)
    \end{aligned}
\end{equation}
with initial condition $x(0)=x_0$.
If the origin is an asymptotically stable equilibrium of \eqref{eq:lpv-cl}, then $x(k)\rightarrow 0$ as $k\rightarrow \infty$ for all possible signals $\theta:\mathbb{N}\rightarrow\Theta$.
Because the system is subject to state and input constraints, this convergence can typically not be attained for all initial conditions $x_0\in\mathbb{X}$.
Therefore, regional asymptotic stability is considered, which is formally defined as follows.

\begin{defn}\label{def:stab-asymp}
    Let $\mathbf{x}(k|\theta,x_0)$ denote the solution $x(k)$ of \eqref{eq:lpv-cl} for a given scheduling signal $\theta:\mathbb{N}\rightarrow\Theta$ and for the initial state $x(0)=x_0$.
    The origin is said to be a regionally asymptotically stable equilibrium of \eqref{eq:lpv-cl},
    if there exists a $\classKL$-function $\beta$ and a proper compact set
    $\mathcal{X}\subseteq\mathbb{X}\subset\mathbb{R}^{n_\mathrm{x}}$ such that
    $\|\mathbf{x}(k|\theta,x_0)\| \leq \beta(\|x_0\|, k)$
    for all possible scheduling signals $\theta:\mathbb{N}\rightarrow\Theta$, for all $x_0\in\mathcal{X}$, and for all $k\in\mathbb{N}$.
\end{defn}

\begin{defn}\label{def:lyap}
    A function $V:\mathbb{R}^{n_\mathrm{x}}\times\mathbb{N}\rightarrow\mathbb{R}$ is a (regional, time-varying) Lyapunov function
    on an invariant proper and compact set $\mathcal{X}\subseteq\mathbb{X}\subset\mathbb{R}^{n_\mathrm{x}}$ for \eqref{eq:lpv-cl} if
    \begin{enumerate}[(i)]
        \item\label{def:lyap:bnd} There exist $\classKinfty$-functions $\lb{v},\ub{v}$ such that for all $(x,k)\in\mathcal{X}\times\mathbb{N}$: $\lb{v}(\|x\|)\leq V(x,k)\leq \ub{v}(\|x\|)$;
        
        \item\label{def:lyap:decr} There exists a $\classK$-function $\delta$ such that for all $(x,\theta,k)\in\mathcal{X}\times\Theta\times\mathbb{N}$:
        $V\left(\Phi\left(x,\theta,k\right), k+1\right) \leq V(x,k) - \delta(\|x\|)$.
    \end{enumerate}
\end{defn}

The next lemma is used to verify the regional asymptotic stability property of Definition~\ref{def:stab-asymp}.

\begin{lem}\cite{Aeyels1998,Jiang2002}\label{lem:lyap}
    If there exists a regional time-varying Lyapunov function satisfying Definition~\ref{def:lyap},
    then the origin is a regionally asymptotically stable equilibrium of \eqref{eq:lpv-cl} in the sense of Definition~\ref{def:stab-asymp}.
\end{lem}


\subsection{Scheduling tubes and anticipative control}
\label{sec:prelim:anticipative}

The value $\theta(k)$ of the scheduling variable can be measured at each time instant $k$.
In principle, for future time instants $k+i$, $i\in\idxo{1}{\infty}$ it is only known that
\begin{equation}\label{eq:theta-unc}
    \theta(k+i)\in\Theta,
\end{equation}
but this assumption can be too restrictive.
In many applications it is known that the scheduling variable can not jump instantaneously over its full range,
but evolves according to a bounded rate-of-variation (ROV) \cite{Casavola2008a,Zheng2013}.
This means that there is a $\delta\theta$ such that for all $k\in\mathbb{N}$,
\begin{equation}\label{eq:theta-rov}
    \left|\theta(k+1) - \theta(k)\right|\leq \delta\theta.
\end{equation}
Thus, the future values $\theta(k+i)$
are known to belong to a ``cone'' expanding outwards from the current point $\theta(k)$.

In some other cases,
the scheduling variable corresponds to a signal that is controlled to follow a reference,
allowing its future evolution to be predicted with high confidence.
This can be described by defining
a nominal signal $\bar{\theta}:\mathbb{N}\rightarrow\Theta$ and an uncertainty $\Delta\subseteq\Theta$ such that
\begin{equation}\label{eq:theta-nom}
    \forall i\in\idxo{0}{\infty}:\ \theta(k+i) \in \left(\bar{\theta}(k+i)\oplus\Delta\right)\cap\Theta.
\end{equation}
If the representation \eqref{eq:lpv-ss} embeds a non-linear system
and its state is controlled to track a reference trajectory,
at each future time instant the state variable belongs to a set $\mathbb{X}_i\subseteq\mathbb{X}$ around this reference.
In an embedding, there is a known relation $\theta=T(x)$ \cite{Hanema2017:cdc:final}.
This gives a description of possible future scheduling trajectories
\begin{equation}\label{eq:theta-nonlin}
    \forall i\in\idxo{0}{\infty}:\ \theta(k+i) \in \left\{T(x)\midsep x\in\mathbb{X}_i\right\}.
\end{equation}
The situations of \eqref{eq:theta-unc}--\eqref{eq:theta-nonlin} represent particular instances of knowledge on possible future trajectories of $\theta$.
To provide a framework in which these and other cases can be described, the notion of ``scheduling tube'' is introduced.

\begin{defn}\label{def:sched-tube}
    A \emph{scheduling tube} $\mathbf{\Theta}$ of length $N$ is a sequence of sets
    $\mathbf{\Theta} = \left\{\Theta_{0},\dots,\Theta_{N-1}\right\} = \seq{\Theta}{0}{N-1}$
    where $\forall i\in\idxcl{0}{N-1}:\Theta_i\subseteq\Theta$, or equivalently, $\mathbf{\Theta}\subseteq\Theta^N$.
\end{defn}

\begin{figure}
    \centering
    \includegraphics[scale=0.9]{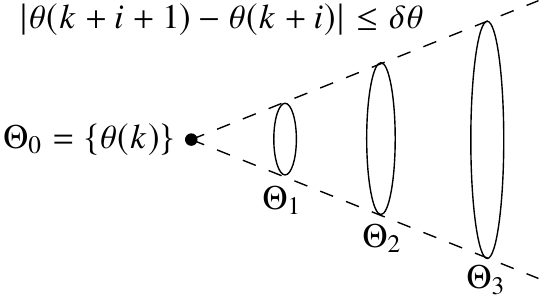}\\
    \includegraphics[scale=0.9]{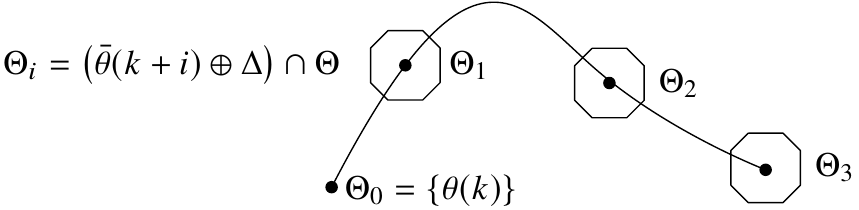}
    \caption{Example of two different scheduling tubes according to cases \protect\eqref{eq:theta-rov} (top) and \protect\eqref{eq:theta-nom} (bottom).\label{fig:anticipative}}
\end{figure}

In the MPC presented in this paper,
at each sampling instant $k$, a new scheduling tube is constructed such that it contains the expected future variation of the scheduling variable.
Due to the availability of the measurement $\theta(k)$,
the scheduling tube is typically constructed such that $\Theta_0=\{\theta(k)\}$.
Then it is assumed that at each instant $k+i$ with $i\in\idxcl{0}{N-1}$, $\theta(k+i)\in \Theta_{i}$ holds.
The sets $\Theta_i$ can be generated using any one of \eqref{eq:theta-unc}--\eqref{eq:theta-nonlin}, or in any other way that fits the application (Figure~\ref{fig:anticipative}).

An operator that can be used to ``order'' scheduling tubes is formally defined next.

\begin{defn}\label{def:order}
    Let $\mathbf{\Theta} = \seq{\Theta}{0}{N-1} \subseteq \Theta^N$ and $\mathbf{\Theta'} = \seq{\Theta'}{0}{N-1} \subseteq \Theta^N$ be two scheduling tubes of length $N$.
    The relation $\mathbf{\Theta}'\sqsubseteq\mathbf{\Theta}$ is satisfied if and only if $\forall i\in\mathbb{N}_{[0,N-2]}:\ \Theta'_i \subseteq \Theta_{i+1}$.
\end{defn}

In Section~\ref{sec:tmpc}, Definition~\ref{def:order} will be used in proving recursive feasibility of the MPC scheme.
In this paper, for computational reasons, it is assumed that all sets in a scheduling tube $\mathbf{\Theta}$ are polytopes.
For notational simplicity it is assumed that these polytopes are all represented as the convex hulls of equally many vertices:

\begin{assum}\label{ass:sched-tube-poly}
    Let $\mathbf{\Theta}\subseteq\Theta^N$ be a scheduling tube according to Definition~\ref{def:sched-tube}.
    Then, every set $\Theta_i$ is a polytope described as the convex hull of $q_\theta$ vertices, i.e.,
    $\forall i\in\idxcl{1}{N-1}:$ $\Theta_i=\mathrm{convh}\{\bar{\theta}^1_i,\dots,\bar{\theta}^{q_\theta}_i\}$.
\end{assum}


\section{HpTMPC: fundamentals}
\label{sec:hpt}

Section~\ref{sec:hpt:intro} introduces the concepts of heterogeneously parameterized tubes (HpTs),
heterogeneous parameterization structure, the tube synthesis problem, and the notion of a domain of attraction (DOA).
Next, the class of cost functions considered in the developed MPC approach is introduced in Section~\ref{sec:hpt:cost}.


\subsection{Parameterized tube synthesis}
\label{sec:hpt:intro}

Before proceeding, a few preliminaries need to be covered.
Define the one-step forward reachable set---or \emph{image}---of a set $X\subset\mathbb{R}^{n_\mathrm{x}}$ for the dynamics \eqref{eq:lpv-ss} under a given controller, and for a corresponding scheduling set, as follows.

\begin{defn}\label{def:image}
    The controlled image of a set for a constrained LPV system represented by the LPV-SS representation
    with a given controller $K:X\times\Theta\rightarrow\mathbb{U}$
    is the map $\img{\cdot}{\cdot}{K}: 2^X\times 2^\Theta \rightarrow 2^{\mathbb{R}^{n_\mathrm{x}}}$ defined by
    $\img{X}{\Theta}{K} = \bigl\{A(\theta)x + B(\theta)K(x,\theta)\mid x\in X,\theta\in\Theta\bigr\}$.
\end{defn}

The following inclusion result is directly implied from Definition~\ref{def:image}:

\begin{lem}\label{lem:image-props}
    Let $K:X\times\Theta\rightarrow\mathbb{U}$.
    For any subset $X'\times\Theta'\subseteq X\times\Theta$, it holds $\img{X'}{\Theta'}{K} \subseteq \img{X}{\Theta}{K}$.
\end{lem}

In what follows, it is useful to consider controllers $K(\cdot,\cdot)$ that satisfy some additional properties.
These properties are summarized here under the name of continuous and positively homogeneous of degree one ($\mathcal{CH}_1$):

\begin{defn}\label{def:conhomogen}
    A controller $K:X\times\Theta\rightarrow\mathbb{U}$ is $\mathcal{CH}_1$ if it is
    (i) a continuous function\footnote{Continuity of $K(\cdot,\cdot)$ ensures that $u=K(x,\theta)$ is well-defined (i.e., single-valued) for all $(x,\theta)$ on its domain $X\times\Theta$.} of its input arguments $(x,\theta)\in X\times \Theta$,
    and (ii) positively homogeneous of degree one in the sense that $\forall \alpha\in\mathbb{R}_+:\ K(\alpha x,\theta)=\alpha K(x,\theta)$.
\end{defn}

Because the representation \eqref{eq:lpv-ss} is also homogeneous, the limitation to $\mathcal{CH}_1$ controllers is not restrictive.
In the definition of a tube, so-called second-order functions will be used to describe parameterized control policies:

\begin{defn}\label{def:second-order-function}
    A function $f:\mathcal{A}\rightarrow\mathcal{B}$ is called first-order if both $\mathcal{A}$ and $\mathcal{B}$ are subsets of real vector spaces.
    The function $f$ is called \emph{second-order} if it returns another first-order function, i.e., if $\mathcal{A}$ is a subset of a real vector space but $\mathcal{B}$ is a subset of all first-order functions $g:\mathcal{C}\rightarrow\mathcal{D}$ (i.e., with $\mathcal{C}$, $\mathcal{D}$ being subsets of real vector spaces).
\end{defn}

Higher-order functions are widely used in computer science as useful abstractions \cite[Chapter~1.3]{Abelson1996}.
A simple second-order function is $g:\mathbb{R}\rightarrow(\mathbb{R}^n\rightarrow\mathbb{R}^n)$, $g(c) = (x\mapsto cx)$.
It is then possible to say, e.g., $h = g(2)$ meaning that $h$ is the function $h:\mathbb{R}^n\rightarrow\mathbb{R}^n$, $h(x)=2x$.

The concept of a tube can now be defined.

\begin{defn}\label{def:tube}
    Let $\mathbf{\Theta}\subseteq\Theta^N$ be given according to Definition~\ref{def:sched-tube}.
    A tube of length $N$ is a pair
    $\mathbf{T} = \bigl(\mathbf{X},\mathbf{K}\bigr) = \left(\seq{X}{0}{N}, \seq{K}{0}{N-1}\right)$
    where $X_i\subseteq\mathbb{R}^{n_{\mathrm{x}}}$ are sets and where $K_i: X_i\times \Theta_i \rightarrow\mathbb{U}$ are $\mathcal{CH}_1$ control laws such that
    for all $i\in\idxcl{0}{N-1}$, the condition $\img{X_i}{\Theta_i}{K_i} \subseteq X_{i+1} \cap \mathbb{X}$ holds.
    Each set $X_i$ is called a cross section.
\end{defn}

The length $N$ of the tube in Definition~\ref{def:tube} is called the \emph{prediction horizon}.
The cross sections $X_i$ are not necessarily subsets of the state constraints.
Requiring this would be conservative, because the cross sections are supersets of the sets of reachable states given by $\img{\cdot}{\cdot}{\cdot}$ \cite{Brunner2013}.
Therefore, in Definition~\ref{def:tube}, only the reachable states given by $\img{\cdot}{\cdot}{\cdot}$ are required to satisfy the state constraints.
The following should also be kept in mind:

\begin{rem}
    It is important to
    remember the difference between a ``tube''
    (a synthesized sequence of sets in the state space with associated controllers, Definition~\ref{def:tube})
    and a ``scheduling tube''
    (a sequence of sets describing possible future values of the scheduling variable, Definition~\ref{def:sched-tube}).
\end{rem}

As all necessary notions have been introduced, a heterogeneously parameterized tube is defined next.

\begin{defn}\label{def:hpt}
    Let $\mathbf{T}$ be a tube according to Definition~\ref{def:tube}.
    For all $i\in\mathbb{N}_{[0,N]}$, introduce parameter sets $\mathbb{P}(i) = \mathbb{P}^\mathrm{x}(i)\times\mathbb{P}^\mathrm{k}(i)$.
    A tube $\mathbf{T}$ is a \emph{heterogeneously parameterized tube (HpT)} if it satisfies:
    \begin{enumerate}[(i)]
        \item For all $i\in\mathbb{N}_{[0,N]}$, there is a set-valued function
            $P^\mathrm{x}\left(\cdot|i\right): \mathbb{P}^\mathrm{x}(i)\rightarrow 2^{\mathbb{R}^{n_\mathrm{x}}}$ and there exists a parameter
            $p^\mathrm{x}_i\in\mathbb{P}^\mathrm{x}(i)$ such that $X_i=P^\mathrm{x}\bigl(p^\mathrm{x}_i|i\bigr)$.

        \item For all $i\in\mathbb{N}_{[0,N-1]}$, there is a second-order function
            $P^\mathrm{k}\left(\cdot|i\right): \mathbb{P}^\mathrm{k}(i)\rightarrow \bigl(X_i\times\Theta_i\rightarrow\mathbb{U}\bigr)$ and there exists a parameter
            $p^\mathrm{k}_i\in\mathbb{P}^\mathrm{k}(i)$ such that $K_i=P^\mathrm{k}\bigl(p^\mathrm{k}_i|i\bigr)$.
    \end{enumerate}
    Furthermore, define the shorthand
    \begin{equation*}
        P(p|i) = \bigl(P^\mathrm{x}\bigl(p^\mathrm{x}|i\bigr), P^\mathrm{k}\bigl(p^\mathrm{k}|i\bigr)\bigr)
    \end{equation*}
    where $p=\bigl(p^\mathrm{x},p^\mathrm{k}\bigr)$.
\end{defn}

The distinguishing feature of the parameterization proposed in the above definition, and the reason why it is called a \emph{heterogeneous} parameterization,
is that the sets $\mathbb{P}(i)$ and functions $P(\cdot|i)$ can be different for every prediction time instant $i\in\mathbb{N}_{[0,N]}$.
If the sets and functions in Definition~\ref{def:hpt} are chosen to be independent of $i$,
the setup of \cite{Hanema2016:cdc:final} (equivalently, the setup of \cite{Hanema2017:aut:final} with $M=1$) is recovered.
With the above definition, a \emph{heterogeneous parameterization structure} can be associated.

\begin{defn}\label{def:pn}
    A heterogeneous parameterization structure $\mathcal{P}_N$ is defined as the sequence of pairs
    \begin{align*}
        \mathcal{P}_N = \bigl\{\left(\mathbb{P}(0), P(\cdot|0)\right),
            \dots,
            \left(\mathbb{P}(N), P(\cdot|N)\right)\bigr\}.
    \end{align*}
\end{defn}

\begin{figure}
    \centering
    \includegraphics[width=\plotwidth]{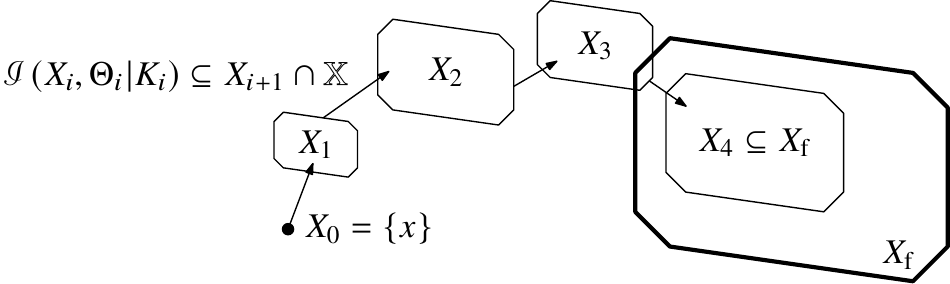}
    \caption{A ``feasible'' tube $\mathbf{T}\in\mathcal{T}_N\left(x,\mathbf{\Theta}|\mathcal{P}_N\right)$ in a two-dimensional state space with $N=4$. Recall that $\mathbf{\Theta}=\seq{\Theta}{0}{N-1}$.\label{fig:feas-tube}}
\end{figure}

The parameterization structure $\mathcal{P}_N$ has to be selected during the control design and determines the computational complexity and the achievable performance of the resulting controller.
In what follows, a tube is called \emph{feasible} if it satisfies an initial condition constraint and a terminal constraint.
Given a structure $\mathcal{P}_N$, the set of such feasible tubes $\mathcal{T}_N(\cdot,\cdot|\mathcal{P}_N)$ can be defined as
\begin{multline}\label{eq:tn}
    \mathcal{T}_N\left(x,\mathbf{\Theta}\nmidsep \mathcal{P}_N\right) =
    \bigl\{\mathbf{T}\,\nmidsep \,\mathbf{T}\ \mathrm{satisfies\ Def.\ \ref{def:hpt}}\ \mathrm{with}\\ X_{0}=\{x\}\ \mathrm{and}\ X_N\subseteq X_\mathrm{f} \bigr\}
\end{multline}
where $X_\mathrm{f}\subseteq\mathbb{X}$ is a terminal set.
An example of a feasible tube is depicted in Figure~\ref{fig:feas-tube}.
Selecting from this set a single tube that optimizes a given performance criterion
is done by solving the \emph{tube synthesis} problem
\begin{equation}\label{eq:tube-synth}
    \begin{aligned}
        V\left(x, \mathbf{\Theta}\nmidsep \mathcal{P}_N\right) &=
        \min_{\mathbf{T}} J_N\left(\mathbf{T},\mathbf{\Theta}\right)\\
        &\ \mathrm{subject\ to}\ \mathbf{T}\in\mathcal{T}_N\left(x, \mathbf{\Theta}\nmidsep \mathcal{P}_N\right),
    \end{aligned}
\end{equation}
where
\begin{equation}\label{eq:cost}
    J_N\left(\mathbf{T}, \mathbf{\Theta}\right) = \sum_{i=0}^{N-1} \ell\left(X_i,K_i,\Theta_i\right) + F\left(X_N\right)
\end{equation}
is a finite-horizon cost function with $\ell(\cdot,\cdot,\cdot)$ being the stage cost and $F(\cdot)$ being the terminal cost.
The terminal cost $F(\cdot)$ must be chosen such that closed-loop stability is guaranteed.
The function $V(\cdot,\cdot|\cdot)$ in \eqref{eq:tube-synth} is called the \emph{value function}, and an optimizer of \eqref{eq:tube-synth} is denoted as
\begin{equation}\label{eq:tstar}
    \mathbf{T}^\star = \left(\seq{X^\star}{0}{N}, \seq{K^\star}{0}{N-1}\right),
\end{equation}
where by definition, $V(x,\mathbf{\Theta}|\mathcal{P}_N) = J_N(\mathbf{T}^\star, \mathbf{\Theta})$.
The DOA is the set of initial states for which a feasible tube $\mathbf{T}\in\mathcal{T}_N(x,\mathbf{\Theta}|\mathcal{P}_N)$ exists, and is formally defined as follows.

\begin{defn}\label{def:doa}
    For a given sequence $\mathbf{\Theta}\subseteq\Theta^N$, the \emph{domain of attraction (DOA)} of the closed-loop system \eqref{eq:lpv-cl} under a controller defined by \eqref{eq:tube-synth} is
    \begin{equation}\label{eq:doa}
        \mathcal{X}_N\left(\mathbf{\Theta}\nmidsep \mathcal{P}_N\right) = \left\{x\in\mathbb{X}\,\nmidsep \,\mathcal{T}_N\left(x,\mathbf{\Theta}\nmidsep \mathcal{P}_N\right)\neq\emptyset\right\}.
    \end{equation}
\end{defn}


\subsection{Cost function design}
\label{sec:hpt:cost}

In this section, the class of stage cost functions used in the developed MPC approach is presented.
Sufficient conditions on the terminal cost $F(\cdot)$ under which the value function $V(\cdot,\cdot|\cdot)$ can be bounded by a pair of \classKinfty-functions are provided.
The norm-based stage cost
\begin{equation}\label{eq:stage-cost}
    \ell\left(X,K,\Theta\right) = \max_{(x,\theta)\in X\times\Theta}\left(\|Qx\|^c + \|RK(x,\theta)\|^c\right)
\end{equation}
is proposed where $\|\cdot\|$ can be any vector norm,
$c\geq 1$,
and $(Q,R)\in\mathbb{R}^{n_\mathrm{q}\times n_\mathrm{x}}\times\mathbb{R}^{n_\mathrm{r}\times n_\mathrm{x}}$ are full column rank matrices corresponding to tuning parameters.
Several properties of \eqref{eq:stage-cost} are important to guarantee stability of the MPC algorithm presented in this paper.
These are summarized in the following proposition.

\begin{prop}\label{prop:stage-cost}
    In what follows, let $K:X\times\Theta\rightarrow\mathbb{U}$ be a $\mathcal{CH}_1$-controller.
    \begin{enumerate}[(i)]
        \item\label{prop:stage-cost:l-subset} For all subsets $X'\times\Theta'\subseteq X\times\Theta$, it holds that
        $\ell\bigl(X',K,\Theta'\bigr) \leq \ell\bigl(X,K,\Theta\bigr)$.

        \item\label{prop:stage-cost:l-bnd} There exists a \classKinfty-function $\lb{\ell}$ such that\\ $\lb{\ell}\bigl(d_H^0\bigl(X\bigr)\bigr)\leq \ell(X, K, \Theta)$.
        
        \item\label{prop:stage-cost:l-homogen} The stage cost is homogeneous of degree $c$ in the sense that for all $\alpha\in\mathbb{R}_+$, $\ell(\alpha X,K,\Theta)=\alpha^c\ell(X,K,\Theta)$.
    \end{enumerate}
\end{prop}
\begin{pf}
    \emph{Proof of (i).}
    This follows from the definition of \eqref{eq:stage-cost} in terms of the maximum over a compact set.
    
    \noindent
    \emph{Proof of (ii).}
    From \eqref{eq:stage-cost}, $\max_{x\in X} \|Qx\|^c \leq \ell\bigl(X,K,\Theta\bigr)$.
    As $Q$ is assumed to be full column rank, $x\mapsto\|Qx\|$ is a norm (in particular, $\|Qx\|=0$ if and only if $x=0$).
    All norms in finite-dimensional vector spaces are equivalent, hence $\exists \alpha>0: \alpha\|x\|\leq\|Qx\|$, implying that $\alpha \max_{x\in X} \|x\| = \alpha d_H^0(X) \leq \max_{x\in X}\|Qx\|$.
    This directly leads to
    $\bigl(\alpha d_H^0(X)\bigr)^c \leq \bigl(\max_{x\in X}\|Qx\|\bigr)^c = \max_{x\in X} \|Qx\|^c$,
    proving the statement with $\lb{\ell}\bigl(\xi\bigr) = \left(\alpha \xi\right)^c$.
    
    \noindent
    \emph{Proof of (iii).}
    This is direct from the $\mathcal{CH}_1$-property of $K(\cdot,\cdot)$ combined with homogeneity of the norm $\|\cdot\|$.
    \qed
\end{pf}

To prove closed-loop stability of a predictive controller, the usual approach is to show that the value function $V(\cdot,\cdot|\cdot)$ of \eqref{eq:tube-synth} is a Lyapunov function in the sense of Definition~\ref{def:lyap}.
An important first step, then, is to show that the value function satisfies Definition~\ref{def:lyap}.(\ref{def:lyap:bnd}), i.e., to show that it can be upper- and lower bounded by a pair of \classKinfty-functions.
The remainder of this section is devoted to proving, under some assumptions, the existence of these bounds.
Besides the stage cost \eqref{eq:stage-cost}, the finite-horizon cost function \eqref{eq:cost} also contains a terminal cost.
An explicit construction of a suitable terminal cost will be given later in Section~\ref{sec:design}.
For now, the following necessary assumptions on the terminal cost are made.

\begin{assum}\label{ass:terminal-cost}
    Consider the terminal cost function $F(\cdot): 2^{\mathbb{R}^{n_\mathrm{x}}}\rightarrow\mathbb{R}_+$ in \eqref{eq:cost}.
    \begin{enumerate}[(i)]
        \item\label{ass:terminal-cost:homogen} Let $c$ have the same value as in \eqref{eq:stage-cost}.
        The function $F(\cdot)$ is homogeneous of degree $c$ in the sense that for all $\alpha\in\mathbb{R}_+$, $F(\alpha X)=\alpha^c F(X)$.
        
        \item\label{ass:terminal-cost:bnd} There exist \classKinfty-functions $\lb{F},\ub{F}$ such that for all $X\subseteq X_\mathrm{f}$, $\lb{F}\bigl(d_H^0\bigl(X\bigr)\bigr)\leq F\bigl(X\bigr)\leq \ub{F}\bigl(d_H^0\bigl(X\bigr)\bigr)$.
    \end{enumerate}
\end{assum}

By Proposition~\ref{prop:stage-cost}.(\ref{prop:stage-cost:l-homogen}) and Assumption~\ref{ass:terminal-cost}.(\ref{ass:terminal-cost:homogen}),
the function $J_N(\cdot,\cdot)$ is homogeneous of degree $c$ in the sense that $\forall \alpha\in\mathbb{R}_+: J_N(\alpha \mathbf{T}, \mathbf{\Theta})=\alpha^c J_N(\mathbf{T}, \mathbf{\Theta})$.
Define the scalar multiple of a tube as $\alpha\mathbf{T} = \bigl(\seq{\alpha X}{0}{N}, \seq{K}{0}{N-1}\bigr)$.
The main result of this section can now be proven.

\begin{prop}\label{prop:value-fun}
    Assume that $X_\mathrm{f}$ in \eqref{eq:tube-synth} is a PC-set.
    Then, there exist \classKinfty-functions $\lb{v},\ub{v}$ such that for all $x\in\mathbb{X}$ and $\mathbf{\Theta}\subseteq\Theta^N$ for which
    \eqref{eq:tube-synth} is feasible, it holds that $\lb{v}\bigl(\|x\|\bigr) \leq V\bigl(x,\mathbf{\Theta}|\mathcal{P}_N\bigr) \leq \ub{v}\bigl(\|x\|\bigr)$.
\end{prop}
\begin{pf}
    The lower bound is established trivially as $V(x,\mathbf{\Theta}|\mathcal{P}_N) \geq \lb{\ell}\left(\|x\|\right)$.
    Let $\partial S$ denote the boundary of a set $S$
    and let $\phi(x) = \psi_{\mathcal{X}_N(\mathbf{\Theta})}(x)$.
    The domain of attraction $\mathcal{X}_N(\mathbf{\Theta})$
    of \eqref{eq:tube-synth} for a given sequence $\mathbf{\Theta}$ is proper and compact.
    Representation \eqref{eq:lpv-ss} is homogeneous of degree one in $(x,u)$, and $(X_\mathrm{f},\mathbb{X},\mathbb{U})$ are PC-sets.
    Therefore, the existence of a
    $\mathbf{T}\in\mathcal{T}_N(x,\mathbf{\Theta}|\mathcal{P}_N)$ implies that for all $\alpha\in[0,1]$,
    there exists a $\mathbf{T}^\circ=\alpha\mathbf{T}\in\mathcal{T}_N(\alpha x,\mathbf{\Theta}|\mathcal{P}_N)$.
    Hence, for any $x\in\mathcal{X}_N(\mathbf{\Theta})$ it holds that $x\in \phi(x) \partial\mathcal{X}_N(\mathbf{\Theta})$, and
    \begin{equation}\label{prop:value-fun:1}
        \begin{aligned}
            V\left(x,\mathbf{\Theta}|\mathcal{P}_N\right)
            &\leq \max_{x\in \phi(x) \partial \mathcal{X}_N\left(\mathbf{\Theta}\right)} V\left(x,\mathbf{\Theta}|\mathcal{P}_N\right)\\
            &= \max_{x\in \partial \mathcal{X}_N\left(\mathbf{\Theta}\right)} V\left(\phi(x)x,\mathbf{\Theta}|\mathcal{P}_N\right).
        \end{aligned}
    \end{equation}
    Because the cost function is homogeneous of degree $c$, it follows that $J_N(\mathbf{T}^\circ, \mathbf{\Theta}) = \alpha^c J_N(\mathbf{T}, \mathbf{\Theta})$.
    Now note that, for all $x\in\mathcal{X}_N\left(\mathbf{\Theta}\right)$, $\phi(x)\in[0,1]$ holds, and that
    the solution $\mathbf{T}^\circ$ is feasible but not necessarily optimal for the initial state $\alpha x$.
    Combining these facts with \eqref{prop:value-fun:1} yields
    \begin{equation}\label{prop:value-fun:2}
        \begin{aligned}
            V\left(x,\mathbf{\Theta}|\mathcal{P}_N\right)
            &\leq \max_{x\in \partial \mathcal{X}_N\left(\mathbf{\Theta}\right)} V\left(\phi(x)x,\mathbf{\Theta}|\mathcal{P}_N\right)\\
            &\leq \phi^c(x) \max_{x\in \partial \mathcal{X}_N\left(\mathbf{\Theta}\right)} V\left(x,\mathbf{\Theta}|\mathcal{P}_N\right).
        \end{aligned}
    \end{equation}
    Because $(X_\mathrm{f},\mathbb{X},\mathbb{U})$ are compact, it can be assumed that there exists a constant $\hat{V}>0$ such that
    \begin{equation}\label{prop:value-fun:3}
        \max_{x\in \partial \mathcal{X}_N\left(\mathbf{\Theta}\right)} V\left(x,\mathbf{\Theta}|\mathcal{P}_N\right) \leq \hat{V}.
    \end{equation}
    As $\phi(\cdot)$ is the gauge function of a proper and compact set,
    there exists a \classKinfty-function $\ub{\phi}$ such that for all $x\in\mathbb{R}^{n_\mathrm{x}}:\ \phi(x)\leq \ub{\phi}(\|x\|)$ \cite[Lemma~1]{Hanema2017:aut:final}.
    Combining this with \eqref{prop:value-fun:2}-\eqref{prop:value-fun:3} gives that
    $\ub{v}(\xi) = (\ub{\phi}(\xi))^c\hat{V}$ is a \classKinfty-upper bound on $V(\cdot,\mathbf{\Theta}|\mathcal{P}_N)$.
    \qed
\end{pf}


\section{HpTMPC: prototype algorithm}
\label{sec:tmpc}

In this section, the HpTs introduced previously are used to construct a stabilizing MPC algorithm.


\subsection{Parameterization conditions}
\label{sec:tmpc:para-conditions}

In this subsection, a number of conditions is presented that allows for the derivation of a recursively feasible and stabilizing MPC algorithm.
These conditions come as a set of assumptions on (i) the existence of a terminal set and local controller, (ii) the parameterization structure $\mathcal{P}_N$, and (iii) the cost function $J_N(\cdot,\cdot)$.
First, two preliminary definitions are provided.

\begin{defn}\label{def:contractive}
    A PC-set $X\subseteq\mathbb{X}$ is called controlled $\lambda$-contractive for an LPV-SS representation \eqref{eq:lpv-ss},
    if there exists a local $\mathcal{CH}_1$-controller $K:X\times\Theta\rightarrow\mathbb{U}$ 
    such that $\lambda = \inf\{\mu\geq 0\nmidsep\img{X}{\Theta}{K}\subseteq \mu X\}<1$.
\end{defn}

Given a controller $K:X\times\Theta\rightarrow\mathbb{U}$, it will turn out to be useful to consider controllers that are ``the same'' as $K$ on a subset of the original domain $X\times\Theta$. 
Formally, the set of \emph{restrictions} of a controller can be defined as follows.

\begin{defn}\label{def:restr}
    Let $K:X\times\Theta\rightarrow\mathbb{U}$.
    The set of restrictions of $K$ to the subset $X'\times\Theta' \subseteq X\times\Theta$ is
    \begin{multline*}
        \restr{K}{X'}{\Theta'} = \bigl\{K':X'\times\Theta'\rightarrow\mathbb{U}\nmidsep\\\forall(x,\theta)\in X'\times\Theta':\ K'(x,\theta) = K(x,\theta)\bigr\}.
    \end{multline*}
\end{defn}

The first set of assumptions required for deriving a recursively feasible and stabilizing MPC can now be stated.

\begin{assum}\label{ass:hptmpc-terminal}
    The terminal set $X_\mathrm{f}$ in \eqref{eq:tube-synth} is controlled $\lambda$-contractive in the sense of Definition~\ref{def:contractive}.
    (Recall that a local controller which renders $X_\mathrm{f}$ $\lambda$-contractive in this sense, is denoted as $K_\mathrm{f}:X_\mathrm{f}\times\Theta\rightarrow\mathbb{U}$.)
\end{assum}

With Assumption~\ref{ass:hptmpc-terminal} in place, the first step towards proving recursive feasibility can be made.
The next lemma on the existence of ``successor tubes'' is required first.

\begin{lem}\label{lem:succ}
    Let $N\in\mathbb{N}_{[1,\infty)}$, let $\mathcal{P}_N$ be a heterogeneous parameterization structure according to Definition~\ref{def:pn}, and
    let $(x,\theta)$ be the current state and scheduling variable values.
    Furthermore, let $\mathbf{\Theta} = \bigl\{\{\theta\}, \seq{\Theta}{1}{N-1}\bigr\} \subseteq \Theta^N$ and $\mathbf{\Theta}^+ = \seq{\Theta^+}{0}{N-1} \subseteq\Theta^N$ be two scheduling tubes satisfying $\mathbf{\Theta}^+ \sqsubseteq \mathbf{\Theta}$.
    Suppose that there exists a tube $\mathbf{T}\in\mathcal{T}_N\bigl(x,\mathbf{\Theta}\nmidsep \mathcal{P}_N\bigr)$.
    Then, there always exists a $\gamma\in[0,1]$ and a sequence of sets $\mathbf{X}^+ = \seq{X^+}{0}{N}$ that satisfies
    \begin{subequations}\label{eq:succ}
        \begin{align}
            \forall i\in\mathbb{N}_{[0,N-2]}:\ & X^+_i \subseteq X_{i+1},\label{eq:succ:x0}\\
            & X^+_{N-1} \subseteq \gamma X_\mathrm{f},\label{eq:succ:xmid}\\
            & X^+_N \subseteq \lambda \gamma X_\mathrm{f},\label{eq:succ:x1}\\
            \forall i\in\mathbb{N}_{[0,N-2]}:\ & \img{X^+_i}{\Theta^+_i}{K_{i+1}} \subseteq X^+_{i+1}\cap\mathbb{X},\label{eq:succ:k0}\\
            & \img{X^+_{N-1}}{\Theta^+_{N-1}}{K_\mathrm{f}} \subseteq X^+_N \label{eq:succ:k1}.
        \end{align}
    \end{subequations}
\end{lem}
\begin{pf}
    By construction of $\mathbf{T}$, $X^+_0 = \{A(\theta)x + BK_0(x,\theta)\}\subseteq X_1$.
    Recall that $\mathbf{\Theta}^+ \sqsubseteq \mathbf{\Theta}$ means that $\Theta^+_i \subseteq \Theta_{i+1}$ for all $i\in\mathbb{N}_{[0,N-2]}$.
    From Lemma~\ref{lem:image-props}, it follows
    $\img{X^+_0}{\Theta^+_0}{K_1} \subseteq \img{X_1}{\Theta_1}{K_1}$.
    Because $\img{X_1}{\Theta_1}{K_1} \subseteq X_2\cap\mathbb{X}$,
    there exists a $X^+_1 \subseteq X_2$ such that $\img{X^+_0}{\Theta^+_0}{K_1} \subseteq X^+_1\cap\mathbb{X}$.
    Repeating this argument for all subsequent time instances $i\in\mathbb{N}_{[1,N-2]}$ yields the existence of a sequence satisfying \eqref{eq:succ:x0} and \eqref{eq:succ:k0}.
    Next, let $\gamma=\inf\{\gamma'\geq 0\,|\,X_N\subseteq\gamma' X_\mathrm{f}\}$.
    By construction, $X_N\subseteq X_\mathrm{f}$, therefore $\gamma\in[0,1]$.
    It was previously shown that $X^+_{N-2}\subseteq X_{N-1}$.
    Thus, $\img{X^+_{N-2}}{\Theta^+_{N-2}}{K_{N-1}} \subseteq X_N \subseteq \gamma X_\mathrm{f}$, giving \eqref{eq:succ:xmid}.
    Homogeneity of \eqref{eq:lpv-ss} combined with Assumption~\ref{ass:hptmpc-terminal} imply that for all $\gamma\in[0,1]$
    the inclusion $\img{\gamma X_\mathrm{f}}{\Theta}{K_\mathrm{f}}\subseteq\lambda\gamma X_\mathrm{f}$ holds.
    Because $X^+_{N-1}\subseteq\gamma X_\mathrm{f}$ and $\Theta^+_{N-1}\subseteq\Theta$, this yields \eqref{eq:succ:x1} and \eqref{eq:succ:k1}, completing the proof.
    \qed
\end{pf}

Lemma~\ref{lem:succ} establishes the existence of a successor tube that satisfies certain properties, but without constructing one.
As apparent from the proof, a sequence $\mathbf{X}^+$ that satisfies the stronger condition $X^+_{N-1} \subseteq X^+_N \subseteq \gamma X_\mathrm{f}$ instead of just \eqref{eq:succ:xmid} exists.
However, for proving recursive feasibility and stability, condition \eqref{eq:succ:xmid} is sufficient and less restrictive in terms of the permissible designs of $\mathcal{P}_N$.

\begin{rem}\label{rem:succ}
    Two particular constructions of the sequence $\mathbf{X^+}$ from Lemma~\ref{lem:succ} are the following.
    First, it is possible to set ``$\subseteq$'' in \eqref{eq:succ:k0}-\eqref{eq:succ:k1} to ``$=$'':
    then, conditions \eqref{eq:succ:x0}-\eqref{eq:succ:x1} are directly implied.
    Second, it is also possible to replace inclusion with equality in \eqref{eq:succ:x0}-\eqref{eq:succ:x1} which, in turn, implies the satisfaction of \eqref{eq:succ:k0}-\eqref{eq:succ:k1}.
    This second option is analogous to ``shifting the sequence'' in standard MPC.
    In this paper, both constructions will be used to prove recursive feasibility of the heterogeneous parameterization that will be proposed in Section~\ref{sec:design:hptsf}.
\end{rem}

Under the condition that a feasible tube $\mathbf{T}$ exists, Lemma~\ref{lem:succ} established the existence of a successor tube $\mathbf{T}^+$,
but without considering if $\mathbf{T}^+$ can be parameterized in the structure $\mathcal{P}_N$.
To ensure that this is the case, the following set of assumptions is invoked.

\begin{assum}\label{ass:hptmpc-rfeas}
    Suppose that the hypotheses of Lemma~\ref{lem:succ} hold true.
    Assume that $\mathcal{P}_N$ is designed such that
    at least one of the possible sequences $\mathbf{X^+}$ fulfilling \eqref{eq:succ} satisfies the following conditions:
    \begin{enumerate}[(i)]
        \item\label{ass:hptmpc-rfeas:pset} For all $i\in\mathbb{N}_{[0,N]}$,
        there exists a $p^{\mathrm{x}+}_i\in\mathbb{P}^\mathrm{x}(i)$ with $P^\mathrm{x}\left(p^{\mathrm{x}+}_i\nmidsep i\right) = X^+_i$.

        \item\label{ass:hptmpc-rfeas:pk} For all $i\in\mathbb{N}_{[1,N-2]}$,
        there exists a $p^{\mathrm{k}+}_i\in\mathbb{P}^\mathrm{k}(i)$ such that $P^\mathrm{k}\left(p^{\mathrm{k}+}_i\nmidsep i\right) \in \restr{K_{i+1}}{X^+_i}{\Theta^+_i}$.
            
        \item\label{ass:hptmpc-rfeas:pkf} There exists a $p^{\mathrm{k}+}_{N-1} \in \mathbb{P}(N-1)$ such that
            $P^\mathrm{k}\left(p^{\mathrm{k}+}_{N-1}\nmidsep N-1\right) \in \restr{K_\mathrm{f}}{X^+_{N-1}}{\Theta^+_{N-1}}$.
    \end{enumerate}
\end{assum}

In Assumption~\ref{ass:hptmpc-rfeas}, it can be argued that conditions (\ref{ass:hptmpc-rfeas:pk})-(\ref{ass:hptmpc-rfeas:pkf}) could be replaced by
$P^\mathrm{k}\left(p^{\mathrm{k}+}_i\nmidsep i\right) = K_{i+1}$ and $P^\mathrm{k}\left(p^{\mathrm{k}+}_{N-1}\nmidsep N-1\right) = K_\mathrm{f}$, respectively.
This is, however, more restrictive.
Suppose for instance that, for some $i$, $X_{i}$ is a set with 2 vertices and $X_{i+1}$ is a structurally different set with 4 vertices
(this exact situation occurs, e.g., in the ``scenario''-parameterization introduced later).
If the corresponding controllers $(K_i,K_{i+1})$ are parameterized as vertex controllers on these sets, then $\mathbb{P}^\mathrm{k}(i)$
is a set in dimension $2n_\mathrm{u}$ whereas $\mathbb{P}^\mathrm{k}(i+1)$ is a set in dimension $4n_\mathrm{u}$.
It is therefore clearly impossible to find a parameter $p^\mathrm{k}_i\in\mathbb{P}(i)$ such that $P^\mathrm{k}\left(p^{\mathrm{k}+}_i\nmidsep i\right) = K^+_i = K_{i+1}$.
In contrast, a parameter $p^\mathrm{k}_i\in\mathbb{P}(i)$ could exist such that $P^\mathrm{k}\left(p^{\mathrm{k}+}_i\nmidsep i\right)$
is a controller $K^+_i$ that produces the same control inputs as $K_{i+1}$ for arguments from the subset $X^+_{i}\subseteq X_{i+1}$.
This is precisely the less restrictive condition which, in Assumption~\ref{ass:hptmpc-rfeas}, is captured formally in terms of sets of restrictions $\mathcal{R}(\cdot|\cdot,\cdot)$.

With Assumption~\ref{ass:hptmpc-rfeas} in place, it is already possible to establish recursive feasibility of an MPC based on \eqref{eq:tube-synth}.
The proof is given together with the proof of closed-loop stability in the proof of Theorem~\ref{thm:props}.
However, for the stability proof, the following set of assumptions on the cost and value function is needed first.

\begin{assum}\label{ass:hptmpc-cost}
    Let $K:X\times\Theta\rightarrow\mathbb{U}$ be a controller
    and assume the following properties:
    \begin{enumerate}[(i)]
        \item\label{ass:hptmpc-cost:f-bnd} There exist $\classKinfty$-functions $\lb{F},\ub{F}$ such that for all $X\subseteq X_\mathrm{f}$, $\lb{F}\bigl(d_H^0\bigl(X\bigr)\bigr)\leq F\bigl(X\bigr)\leq \ub{F}\bigl(d_H^0\bigl(X\bigr)\bigr)$.
        
        \item\label{ass:hptmpc-cost:f-decr} Let $K_\mathrm{f}: X_\mathrm{f}\times\Theta\rightarrow\mathbb{U}$ be the local controller from Assumption~\ref{ass:hptmpc-terminal}.
            Then, for all $X\subseteq X_\mathrm{f}$, it holds that $F\bigl(\img{X}{\Theta}{K_\mathrm{f}}\bigr) - F\bigl(X\bigr) \leq -\ell\bigl(X, K_\mathrm{f}, \Theta\bigr)$.
            
        \item\label{ass:hptmpc-cost:f-scl} For any set $X\subseteq X_\mathrm{f}$ and any subset $X'\subseteq X$, $F(X')\leq F(X)$.
        Further, let $\gamma\in[0,1]$ be the infimal $\gamma$ such that $X\subseteq \gamma X_\mathrm{f}$. Then, $F(X)=F(\gamma X_\mathrm{f})$.
            
        \item\label{ass:hptmpc-cost:v-bnd} There exist $\classKinfty$-functions $\lb{v},\ub{v}$ such that for all $x\in\mathbb{R}^{n_\mathrm{x}}$ and $\mathbf{\Theta}\subseteq\Theta^N$ for which
            \eqref{eq:tube-synth} is feasible, it holds $\lb{v}\bigl(\|x\|\bigr) \leq V\bigl(x,\mathbf{\Theta}\nmidsep \mathcal{P}_N\bigr) \leq \ub{v}\bigl(\|x\|\bigr)$.
    \end{enumerate}
\end{assum}

Assumptions~\ref{ass:hptmpc-cost}.(\ref{ass:hptmpc-cost:f-bnd})-(\ref{ass:hptmpc-cost:f-decr}) can be seen as
``set-based versions'' of the standard set of assumptions on the terminal set and cost in nominal MPC \cite{Mayne2000}.
A construction for the terminal cost is provided, together with an implementable tube parameterization that satisfies Assumption~\ref{ass:hptmpc-rfeas}, in Section~\ref{sec:design}.
From Definition~\ref{def:restr} and Proposition~\ref{prop:stage-cost}.(\ref{prop:stage-cost:l-subset}),
the next result follows:

\begin{cor}\label{cor:l-subset}
    For all subsets $X'\times\Theta'\subseteq X\times\Theta$, it holds that
    $\ell\bigl(X',K,\Theta'\bigr) = \ell\bigl(X',K',\Theta'\bigr) \leq \ell\bigl(X,K,\Theta\bigr)$ for any $K'\in\mathcal{R}\bigl(K|X',\Theta'\bigr)$.
\end{cor}


\subsection{Main result}
\label{sec:tmpc:main-result}

The complete receding-horizon heterogeneously parameterized TMPC algorithm is given in Algorithm~\ref{alg:rhc}.
In Step~\ref{alg:rhc:sched-tube} of the algorithm, a scheduling tube containing the possible future trajectories of the scheduling variables is constructed.
Examples of possible constructions were described in Section~\ref{sec:prelim:anticipative}.
In the next theorem, the main properties of Algorithm~\ref{alg:rhc} are proven.
For brevity and consistency with previous notation, the time indices $k$ shown in Algorithm~\ref{alg:rhc} are not explicitly written in the theorem:
the notation $(\mathbf{\Theta},\mathbf{\Theta}^+)$ corresponds to $(\mathbf{\Theta}_k,\mathbf{\Theta}_{k+1})$ in the algorithm.

\begin{algorithm}
    \algnewcommand\algorithmicassume{\textbf{Assume:}}
    \algnewcommand\Assume{\item[\algorithmicassume]}
    \begin{algorithmic}[1]
        \Require $N\in\mathbb{N}_{[1,\infty)}$ and a structure $\mathcal{P}_N$.
        \Assume Step~5 is feasible at $k=0$, i.e., $\mathcal{T}_N\left(x(0),\mathbf{\Theta}_0\nmidsep \mathcal{P}_N\right)\neq\emptyset$.
        \State $\mathbf{\Theta}_{-1}\gets \Theta^N$
        \State $k\gets 0$
        \Loop
            \State Construct $\mathbf{\Theta}_k = \left\{\{\theta(k)\}, \seqk{\Theta}{1}{N-1}\right\}\subseteq\Theta^N$ such that $\mathbf{\Theta}_k\sqsubseteq\mathbf{\Theta}_{k-1}$\label{alg:rhc:sched-tube}
            \State Solve \eqref{eq:tube-synth} to obtain $\mathbf{T}^\star\in\mathcal{T}_N\left(x(k),\mathbf{\Theta}_k\nmidsep \mathcal{P}_N\right)$\label{alg:rhc:tn2}
            \State Apply $u(k) = K_{0}^\star\left(x(k), \theta(k)\right) = u^\star_{0}$ to \eqref{eq:lpv-ss}
            \State $k\gets k+1$
        \EndLoop
    \end{algorithmic}
    \caption{The receding-horizon HpTMPC algorithm.\label{alg:rhc}}
\end{algorithm}

\begin{thm}\label{thm:props}
    Let the hypotheses of Lemma~\ref{lem:succ} and Assumptions \ref{ass:hptmpc-rfeas}--\ref{ass:hptmpc-cost} be satisfied.
    Let $\mathcal{X}_N(\mathbf{\Theta})$ be the domain of attraction such that for all $x\in\mathcal{X}_N(\mathbf{\Theta})$,
    there exists a solution $\mathbf{T}^\star = \bigl(\seq{X^\star}{0}{N}, \seq{K^\star}{0}{N-1}\bigr) \in \mathcal{T}_N\bigl(x, \mathbf{\Theta}\nmidsep \mathcal{P}_N\bigr)$.
    Then, if $x(k)\in \mathcal{X}_N(\mathbf{\Theta})$, Algorithm~\ref{alg:rhc} achieves the following properties:
    \begin{enumerate}[(i)]
        \item After applying $u(k) = K^\star_0\left(x(k),\theta(k)\right)$ to the system, there exists a guaranteed feasible solution $\mathbf{T}^+ \in \mathcal{T}_N\bigl(x(k+1), \mathbf{\Theta}^+\nmidsep \mathcal{P}_N\bigr)$.
        
        \item The system \eqref{eq:lpv-ss} in closed-loop with the controller defined by \eqref{eq:tube-synth} is regionally asymptotically stable on $\mathcal{X}_N(\mathbf{\Theta})$.
    \end{enumerate}
\end{thm}
\begin{pf}
    \emph{Proof of (i).}
    From Definition~\ref{def:restr}, it follows that
    for any controller $K:X\times\Theta\rightarrow\mathbb{U}$ and for arbitrary subsets $(X'\times\Theta') \subseteq (X\times\Theta)$, we have
    that for all $K'\in\restr{K}{X'}{\Theta'}$ there holds $\img{X'}{\Theta'}{K'} = \img{X'}{\Theta'}{K}$.
    Thus, with the sequence $\mathbf{X}^+=\seq{X^+}{0}{N}$ that is shown to exist in Lemma~\ref{lem:succ}, it is possible to associate a sequence of restricted controllers
    $\mathbf{K}^+ = \seq{K^+}{0}{N-1}$ where
    \begin{align*}
        \forall i\in\mathbb{N}_{[0,N-2]}:\ &K^+_i \in \restr{K^\star_{i+1}}{X^+_{i}}{\Theta^+_{i}},\\
        & K^+_{N-1} \in \restr{K_\mathrm{f}}{X^+_{N-1}}{\Theta^+_{N-1}}.
    \end{align*}
    Then,
    $\mathbf{T}^+ = \left(\mathbf{X}^+, \mathbf{K}^+\right)$
    satisfies the initial condition and terminal constraints and Condition~(i) of Definition~\ref{def:hpt}.
    By Assumption~\ref{ass:hptmpc-rfeas}, there exists a selection of $\mathbf{X}^+$ such that, under the given parameterization structure $\mathcal{P}_N$, conditions (ii)-(iii) of Definition~\ref{def:hpt} are also satisfied.
    Therefore, there exists a $\mathbf{T}^+ \in \mathcal{T}_N\bigl(x(k+1),\mathbf{\Theta}^+\nmidsep \mathcal{P}_N\bigr)$.
    
    
    \noindent
    \emph{Proof of (ii).}
    The standard approach to show that the value function of \eqref{eq:tube-synth} is a Lyapunov function for the closed-loop system is used.
    Recall that $\mathbf{T}^\star=\bigl(\seq{X^\star}{0}{N}, \seq{K^\star}{0}{N-1}\bigr) \in \mathcal{T}_N\bigl(x(k), \mathbf{\Theta}\nmidsep \mathcal{P}_N\bigr)$ 
    was the tube synthesized at the initial time instant $k$.
    Introduce $\Delta V_k = V\bigl(x(k+1), \mathbf{\Theta}^+\nmidsep \mathcal{P}_N\bigr) - V\bigl(x(k),\mathbf{\Theta}\nmidsep \mathcal{P}_N\bigr)$.
    Then
    \begin{multline}\label{eq:props:ineq1}
            \Delta V_k \leq J_N\left(\mathbf{T}^+, \mathbf{\Theta}^+\right) - J_N\left(\mathbf{T}^\star, \mathbf{\Theta}\right)\\
            = \sum_{i=0}^{N-1} \ell\left(X^+_i, K^+_i, \Theta^+_i\right) - \sum_{i=0}^{N-1} \ell\left(X^\star_i, K^\star_i, \Theta_i\right)\\
            + F\left(X^+_N\right) - F\left(X^\star_N\right).
    \end{multline}
    Using \eqref{eq:succ:x0}-\eqref{eq:succ:x1}, $\mathbf{\Theta}^+\sqsubseteq\mathbf{\Theta}$, and Proposition~\ref{prop:stage-cost}.(\ref{prop:stage-cost:l-subset}):
    \begin{multline*}
        \sum_{i=0}^{N-1} \ell\left(X^+_i, K^+_i, \Theta^+_i\right)
        \leq \sum_{i=1}^{N-1} \ell\left(X^\star_i, K^\star_i, \Theta_i\right)\\
        + \ell\left(X_{N-1}^+, K_{N-1}^+, \Theta_{N-1}^+\right).
    \end{multline*}
    Substituting this in \eqref{eq:props:ineq1} yields
    \begin{multline}\label{eq:props:ineq2}
            \!\!
            \Delta V_k
            \leq \sum_{i=1}^{N-1} \ell\left(X^\star_i, K^\star_i, \Theta_i\right) + \ell\left(X^+_{N-1}, K^+_{N-1}, \Theta^+_{N-1}\right)\\
                - \sum_{i=0}^{N-1} \ell\left(X^\star_i, K^\star_i, \Theta_i\right)
                + F\left(X^+_N\right) - F\left(X^\star_N\right)\\
            = -\ell\left(X^\star_0, K^\star_0, \Theta_0\right) + \ell\left(X^+_{N-1}, K^+_{N-1}, \Theta^+_{N-1}\right)\\
                + F\left(X^+_N\right) - F\left(X^\star_N\right).
   \end{multline}
   It is known that both $X^+_{N-1}\subseteq \gamma X_\mathrm{f}$ and $X_N^\star\subseteq\gamma X_\mathrm{f}$, so
   $K^+_{N-1}\in\mathcal{R}(K_\mathrm{f}|X^+_{N-1},\Theta^+_{N-1})\subseteq\mathcal{R}(K_\mathrm{f}|\gamma X_\mathrm{f},\Theta^+_{N-1})$.
   Invoking Proposition~\ref{prop:stage-cost}.(\ref{prop:stage-cost:l-subset}) then gives
   \begin{equation*}
        \ell(X^+_{N-1}, K^+_{N-1}, \Theta^+_{N-1})\leq\ell(\gamma X_\mathrm{f}, K_\mathrm{f}, \Theta^+_{N-1}).
   \end{equation*}
   Assumption~\ref{ass:hptmpc-cost}.(\ref{ass:hptmpc-cost:f-scl}) yields $F(X^\star_N)=F(\gamma X_\mathrm{f})$ and
   since $X^+_N \subseteq \lambda\gamma X_\mathrm{f}$, also
   $F(X^+_N) \leq F(\lambda\gamma X_\mathrm{f}) = F(\img{\gamma X_\mathrm{f}}{\Theta}{K_\mathrm{f}})$.
   Making the appropriate substitutions in \eqref{eq:props:ineq2} leads to
   \begin{multline*}
        \Delta V_k
        \leq -\ell\left(X^\star_0, K^\star_0, \Theta_0\right) + \ell\left(\gamma X_\mathrm{f}, K_\mathrm{f}, \Theta^+_{N-1}\right)\\
             + F\left(\img{\gamma X_\mathrm{f}}{\Theta}{K_\mathrm{f}}\right) - F\left(\gamma X_\mathrm{f}\right),
    \end{multline*}
    and using that $\Theta^+_{N-1}\subseteq\Theta$ subsequently gives
    \begin{multline*}
        \Delta V_k
        \leq -\ell\left(X^\star_0, K^\star_0, \Theta_0\right) + \ell\left(\gamma X_\mathrm{f}, K_\mathrm{f}, \Theta\right)\\
             + F\left(\img{\gamma X_\mathrm{f}}{\Theta}{K_\mathrm{f}}\right) - F\left(\gamma X_\mathrm{f}\right).
    \end{multline*}
    From Proposition~\ref{prop:stage-cost}.(\ref{prop:stage-cost:l-bnd}) and Assumption~\ref{ass:hptmpc-cost}.(\ref{ass:hptmpc-cost:f-decr}) it follows finally that
    \begin{multline*}
        \Delta V_k
            \leq -\ell\left(X^\star_0, K^\star_0, \Theta_0\right)\\
            \leq -\lb{\ell}\left(d_H^0\left(X^\star_0\right)\right)
            = -\lb{\ell}\left(\|x(k)\|\right).
    \end{multline*}
    This, with Assumption~\ref{ass:hptmpc-cost}.(\ref{ass:hptmpc-cost:v-bnd}),
    is sufficient to conclude that $V(\cdot,\cdot|\mathcal{P}_N)$ is a regional Lyapunov function on $\mathcal{X}_N(\mathbf{\Theta})$ in the sense of Definition~\ref{def:lyap}.
    By Lemma~\ref{lem:lyap}, this implies that the origin is a regionally asymptotically stable equilibrium of the closed-loop system.
    \qed
\end{pf}


\section{HpTMPC: tractable parameterizations}
\label{sec:design}

In this section, implementable constructions of a terminal cost and tube parameterization are proposed,
that satisfy the conditions under which the HpTMPC approach was shown to be recursively feasible and asymptotically stabilizing in Section~\ref{sec:tmpc}.


\subsection{The HpT terminal cost}
\label{sec:design:cost}

This subsection proposes a terminal cost function such that Assumption~\ref{ass:hptmpc-cost} is satisfied, giving a stabilizing MPC according to Theorem~\ref{thm:props}.
Using Definition~\ref{def:set-gauge},
the terminal cost is defined in terms of the set-gauge $\Psi_{X_\mathrm{f}}(\cdot)$ as
\begin{equation}\label{eq:terminal-cost}
    F\left(X\right) = \frac{\bar{\ell}}{1-\lambda^c} \Psi_{X_\mathrm{f}}^c\left(X\right),
\end{equation}
where $c$ has the same value as in \eqref{eq:stage-cost}, $X_\mathrm{f}$ and $\lambda$ are according to Assumption~\ref{ass:hptmpc-terminal}, and where
\begin{equation}\label{eq:lbar}
    \bar{\ell} = \ell\left(X_\mathrm{f}, K_\mathrm{f}, \Theta\right)
\end{equation}
is a constant.
In the next proposition, it is shown that the cost functions \eqref{eq:stage-cost} and \eqref{eq:terminal-cost} satisfy Assumption~\ref{ass:hptmpc-cost},
and therefore lead to a stable closed-loop system.

\begin{prop}\label{prop:cost}
    The stage and terminal cost \eqref{eq:terminal-cost} together satisfy Assumptions~\ref{ass:hptmpc-cost}.(\ref{ass:hptmpc-cost:f-bnd})-(\ref{ass:hptmpc-cost:f-scl}).
    Furthermore, the value function $V(\cdot,\cdot|\mathcal{P}_N)$ is $\classKinfty$-bounded in the sense of Assumption~\ref{ass:hptmpc-cost}.(\ref{ass:hptmpc-cost:v-bnd}).
\end{prop}
\begin{pf}
    \emph{Satisfaction of Assumption~\ref{ass:hptmpc-cost}.(\ref{ass:hptmpc-cost:f-bnd}).}
    Because \eqref{eq:terminal-cost} is simply the set-gauge function of the PC-set $X_\mathrm{f}$ raised to the power $c\geq 1$ and multiplied with a constant scalar factor, this property follows directly from the existence of the bounds $\lb{s}_\psi,\ub{s}_\psi$ stated in Lemma~1 of \cite{Hanema2017:aut:final}.
    
    \noindent
    \emph{Satisfaction of Assumption~\ref{ass:hptmpc-cost}.(\ref{ass:hptmpc-cost:f-decr}).}
    By $\lambda$-contractivity of $X_\mathrm{f}$ (Assumption~\ref{ass:hptmpc-terminal}.(i)),
    $\Psi_{X_\mathrm{f}}\bigl(\img{X}{\Theta}{K_\mathrm{f}}\bigr) \leq \lambda \Psi_{X_\mathrm{f}}\bigl(X\bigr)$.
    Also, because of the homogeneity of $K_\mathrm{f}$ (Assumption~\ref{ass:hptmpc-terminal}.(ii)), the stage cost \eqref{eq:stage-cost} is positively homogeneous of degree $c$ in the sense that for any $X\subseteq X_\mathrm{f}$,
    \begin{align*}
        \ell\left(X, K_\mathrm{f}, \Theta\right)
        &\leq \ell\left(\Psi_{X_\mathrm{f}}\left(X\right) X_\mathrm{f}, K_\mathrm{f}, \Theta\right)\\
        &= \Psi_{X_\mathrm{f}}^c\left(X\right) \underbrace{\ell\left(X_\mathrm{f}, K_\mathrm{f}, \Theta\right)}_{\bar{\ell}}.
    \end{align*}
    Thus we have the inequality
    \begin{multline*}
        F\left(\img{X}{\Theta}{K_\mathrm{f}}\right) - F\left(X\right)\\
        = \frac{\bar{\ell}}{1-\lambda^c} \left(\Psi_{X_\mathrm{f}}^c\left(\img{X}{\Theta}{K_\mathrm{f}}\right)- \Psi_{X_\mathrm{f}}^c\left(X\right)\right)\\
        \leq \frac{\bar{\ell}}{1-\lambda^c} \left(\lambda^c \Psi_{X_\mathrm{f}}^c\left(X\right) - \Psi_{X_\mathrm{f}}^c\left(X\right)\right)\\
        = -\bar{\ell} \Psi_{X_\mathrm{f}}^c\left(X\right)
        \leq -\ell\left(X,K_\mathrm{f},\Theta\right).
    \end{multline*}
    
    \noindent
    \emph{Satisfaction of Assumption~\ref{ass:hptmpc-cost}.(\ref{ass:hptmpc-cost:f-scl}).}
    This property is immediate from the definition of $\Psi_{X_\mathrm{f}}(\cdot)$, see Definition~\ref{def:set-gauge}.
    
    \noindent
    \emph{Satisfaction of Assumption~\ref{ass:hptmpc-cost}.(\ref{ass:hptmpc-cost:v-bnd}).}
    Under the three properties proven above, this property has already been proven in Proposition~\ref{prop:value-fun}.
    \qed
\end{pf}


\subsection{The HpT-SF parameterization}
\label{sec:design:hptsf}

In this subsection, one possible design of the parameterization structure $\mathcal{P}_N$ that satisfies Assumption~\ref{ass:hptmpc-rfeas} is proposed.
As a preliminary, the following definition of convex multipliers is introduced.
This can be used to compactly describe vertex control laws later.

\begin{defn}\label{def:convm}
    Let $W=\mathrm{convh}\left\{\bar{w}^1,\dots,\bar{w}^{q_\mathrm{w}}\right\}\subset \mathbb{R}^n$.
    Then $\mathrm{convm}\left(\cdot|W\right): W\rightarrow\mathbb{R}^{q_\mathrm{w}}_+$ is the function
    \begin{multline*}
        \mathrm{convm}\left(w|W\right)\\
        = \arg\inf_{\eta\in\mathbb{R}^{q_\mathrm{w}}_+}\left\{\|\eta\| \midsep \sum_{i=1}^{q_\mathrm{w}} \eta_i\bar{w}^i = w, \sum_{i=1}^{q_\mathrm{w}} \eta_i = 1\right\}.
    \end{multline*}
\end{defn}

In general the multipliers $\eta$ are non-unique, but in Definition~\ref{def:convm} a unique choice is made by minimizing the norm $\|\eta\|$.
Define a ``vertex control'' policy as follows:

\begin{defn}\label{def:vertpol}
    Let $X=\mathrm{convh}\left\{\bar{x}^1,\dots,\bar{x}^{q_\mathrm{x}}\right\}\subset\mathbb{R}^{n_\mathrm{x}}$ and,
    likewise, let $\Theta=\mathrm{convh}\left\{\bar{\theta}^1,\dots,\bar{\theta}^{q_\theta}\right\}$.
    Let
    \begin{equation*}
        p^\mathrm{k}=\left(u^{(1,1)},\dots,u^{(q_\mathrm{x},1)},\dots,u^{(q_\mathrm{x},q_\theta)}\right)\in \mathbb{U}^{q_\theta q_\mathrm{x}}
    \end{equation*}
    be a list of corresponding control actions.
    Then $\mathrm{vertpol}\left(\cdot | X, \Theta\right): \mathbb{U}^{q_\theta q_\mathrm{x}}\rightarrow\left(X\times\Theta\rightarrow \mathbb{U}\right)$
    is a second-order function defined such that
    \begin{multline*}
        \mathrm{vertpol}\left(p^\mathrm{k}\nmidsep X, \Theta\right)\left(x,\theta\right)\\
        = \sum_{i=1}^{q_\mathrm{x}}\left[\mathrm{convm}\left(x | X\right)\right]_i\sum_{j=1}^{q_\theta}\left[\mathrm{convm}\left(\theta|\Theta\right)\right]_j u^{(i,j)},
    \end{multline*}
    where $[v]_i$ denotes the $i$-th element of a vector $v$.
\end{defn}

To develop an implementable parameterization that leads to a convex finite-dimensional optimization problem \eqref{eq:tube-synth},
non-convex constraints must not arise due to the multiplication of a parameter-dependent input matrix with a controller
(i.e., a decision variable) which is dependent on the same scheduling parameter.
This is guaranteed by the following assumption.

\begin{assum}\label{ass:implementable}
    Assume that for all $k$, every set in the scheduling tube constructed in Step~\ref{alg:rhc:sched-tube} of Algorithm~\ref{alg:rhc} is a polytope.
    Denote $K_i=P^\mathrm{k}\bigl(p^\mathrm{k}_i|i\bigr)$ for some $p^\mathrm{k}_i\in\mathbb{P}(i)$.
    Let the elements of $\Theta$ be partitioned as
    \begin{multline*}
        \theta = \begin{bmatrix}\tilde{\theta}_1\\ \tilde{\theta}_2\end{bmatrix},\
        \mathrm{where}\
        \tilde{\theta}_1\in\tilde{\Theta}_1\subseteq\mathbb{R}^{n_{\theta 1}},\
        \tilde{\theta}_2\in\tilde{\Theta}_2\subseteq\mathbb{R}^{n_{\theta 2}},
    \end{multline*}
    where $n_{\theta 1}+n_{\theta 2}=n_{\theta}$,
    and with $\tilde{\Theta}_1$ and $\tilde{\Theta}_2$ being the projections of $\Theta$ onto the first $n_{\theta 1}$ and last $n_{\theta 2}$ dimensions respectively.
    Assume that $\mathcal{P}_N$ is such that for all $i\in\mathbb{N}_{[1,N-1]}$ and for all $\bigl(p^\mathrm{k}_i,\theta\bigr)\in\mathbb{P}(i)\times\Theta$ the product $B(\theta)K_i(x,\theta)$ equals $B(\theta)K_i(x,\theta)=B(\tilde{\theta}_1)K_i(x,\tilde{\theta}_2)$.
\end{assum}

Assumption~\ref{ass:implementable} requires that if the input matrix $B(\cdot)$ of the LPV representation depends on a scheduling variable $\theta_1$,
then the synthesized controllers are independent of $\theta_1$.
They can, however, depend on any other scheduling variable $\theta_2$ that is not related to $B(\cdot)$.

\begin{rem}\label{rem:implementable}
    Two ``extreme'' possibilities that guarantee satisfaction of Assumption~\ref{ass:implementable} are the following:
    \begin{enumerate}[(i)]
        \item $B(\cdot)$ in \eqref{eq:lpv-ss} is constant, i.e., $\forall \theta\in\Theta:\ B(\theta)=B$.
        
        \item The structure $\mathcal{P}_N$ is such that the controllers $K_i:X_i\times\Theta_i\rightarrow\mathbb{U}$ are independent of $\theta$.
    \end{enumerate}
\end{rem}

If $B(\cdot)$ is dependent on the scheduling variables $\theta_1$,
and if it is not possible to make the controller parameterization independent of $\theta_1$,
Assumption~\ref{ass:implementable} can not be satisfied.
In that case, it is an option to expand the system \eqref{eq:lpv-ss} by adding a pre-filter which makes the input matrix of the expanded system parameter-independent \cite{Blanchini2007}.

Under Assumption~\ref{ass:implementable}, a so-called \emph{full scenario tube} can be described in the HpT framework of Definition~\ref{def:hpt}.

\begin{defn}[HpT-S]\label{def:hpts}
    A tube $\mathbf{T}\in\mathcal{T}_N(x,\mathbf{\Theta}|\mathcal{P}_N)$ is a scenario tube or HpT-S if
    $\forall i\in\mathbb{N}_{[0,N-1]}: X_{i+1} = \img{X_i}{\Theta_i}{K_i}$
    and each $K_i$ is parameterized as a vertex controller on the set $X_i\times\Theta_i$.
    Let $q_\mathrm{x}(i)=q_\theta^{\max\{0,i-1\}}$ and $q_\mathrm{k}(i)=q_\theta^{i}$.
    Equivalently, in terms of Definition~\ref{def:hpt} and under Assumption~\ref{ass:implementable},
    a tube is called an HpT-S if the parameterization structure $\mathcal{P}_N$ satisfies
    \begin{enumerate}[(i)]
        \item For all $i\in\mathbb{N}_{[0,N]}$, it holds that $\mathbb{P}^\mathrm{x}(i) = \mathbb{R}^{n_\mathrm{x}q_\mathrm{x}(i)}$, $p^\mathrm{x}_i = \left(\bar{x}^{1}_i,\dots,\bar{x}^{q_\mathrm{x}(i)}_i\right)$,
        and $P^\mathrm{x}\left(p^\mathrm{x}\nmidsep i\right) = \mathrm{convh}\left\{p^\mathrm{x}\right\}$.
        
        \item For all $i\in\mathbb{N}_{[0,N-1]}$, it holds that $\mathbb{P}^\mathrm{k}(i) = \mathbb{R}^{n_\mathrm{u}q_\mathrm{k}(i)}$, $p^\mathrm{k}_i = \left(\bar{u}^{1}_i,\dots,\bar{u}^{q_\mathrm{k}(i)}_i\right)$,
        and $P^\mathrm{k}\left(p^\mathrm{k}|i\right) = \mathrm{vertpol}\left(p^\mathrm{k}\nmidsep X_i,\Theta_i\right)$.
    \end{enumerate}
\end{defn}

Scenario tubes as
alternative solutions for the min-max feedback control problem were proposed in
\cite{Lucia2013,Maiworm2015} for non-linear systems subject to discrete-valued disturbances,
for LTI systems subject to additive disturbances in \cite{Scokaert1998,Kerrigan2004},
and for general uncertain linear systems in \cite{MunozDeLaPena2005,MunozDeLaPena2006}.
An HpT-S is non-conservative, but the number of vertices of the sets $X_{i}$ increases exponentially as each set has $q_\theta$ times more vertices than the preceding set.
In \cite{Lucia2013,Maiworm2015,Lucia2014}, it is proposed to avoid this growth by assuming that the uncertainty resolves after $N_0$ prediction steps,
at which point the tree stops branching.
If this assumption is not met in reality, the scheme loses its feasibility and stability properties.
Here the exponential growth is avoided differently: namely, by switching from a scenario parameterization to a parameterization with fixed-complexity cross sections after $N_0$ prediction steps.
In this way, feasibility and stability guarantees are retained under a more realistic handling of the uncertainty.
The considered fixed-complexity parameterization is defined next.

\begin{defn}[HpT-F]\label{def:hptf}
    A tube $\mathbf{T}\in\mathcal{T}_N(x,\mathbf{\Theta}|\mathcal{P}_N)$ is an HpT-F, if for all $i\in\mathbb{N}_{[0,N-1]}$, $X_i = z_i\oplus \alpha_i X_\mathrm{f}$ where
    $\alpha_i\in\mathbb{R}_+$ is the corresponding \emph{cross-section scaling} and $z_i\in\mathbb{R}^{n_\mathrm{x}}$ is the \emph{cross-section center}.
    Equivalently, in terms of Definition~\ref{def:hpt}, a tube is called an HpT-F if the parameterization structure $\mathcal{P}_N$ satisfies
    \begin{enumerate}[(i)]
        \item For all $i\in\mathbb{N}_{[0,N]}$, it holds that $\mathbb{P}^\mathrm{x}(i) = \mathbb{R}^{1+n_\mathrm{x}}$, $p^\mathrm{x}_i=(z_i, \alpha_i)$,
            and $P^\mathrm{x}\bigl(p^\mathrm{x}\nmidsep i\bigr) = z\oplus\alpha X_\mathrm{f}$.
            
        \item For $i\in\mathbb{N}_{[1,N-1]}$, the sets $\mathbb{P}^\mathrm{k}(i)$ and functions $P^\mathrm{k}(\cdot|i)$ are such that
            if $\exists p^\mathrm{k}_{i}\in\mathbb{P}^\mathrm{k}(i)$ with $P^\mathrm{k}\bigl(p^\mathrm{k}_{i}\nmidsep i\bigr) = K$, then
            $\exists p^\mathrm{k}_{i-1}\in\mathbb{P}^\mathrm{k}(i-1)$ satisfying $P^\mathrm{k}\bigl(p^\mathrm{k}_{i-1}\nmidsep i-1\bigr) = K$.
    \end{enumerate}
\end{defn}

In the HpT-F, as the cross-sections $X_i$ are scaled and translated versions of the same set $X_\mathrm{f}$, it is said that the cross sections are \emph{homothetic} to $X_\mathrm{f}$.
The parameterization of the associated control laws is still allowed to be time-varying along the prediction horizon,
i.e., the sets and functions $(\mathbb{P}^\mathrm{k}(\cdot), P(\cdot|\cdot))$ are dependent on $i$.
Thus, an HpT-F satisfying \ref{def:hptf} is still called ``heterogeneous''.
The condition of Definition~\ref{def:hptf}.(ii) is required to ensure that an HpT-F satisfies Assumption~\ref{ass:hptmpc-rfeas}.(ii)-(iii),
so that it leads to a recursively feasible MPC.
In Table~\ref{tab:paras}, some control parameterizations are listed for illustration.

\begin{table}
    \centering
    \begin{tabular}{llcl}
        & Parameterization & DOF\\\hline
        1 & $K_i(x,\theta) = c_i + K_\mathrm{f}(x-z_i,\theta)$ & $1$\\
        2 & $K_i(x,\theta) = c_i(\theta) + K_\mathrm{f}(x-z_i,\theta)$ & $q_\theta$\\
        3 & $K_i(x,\theta) = \mathrm{vertpol}\left(p^\mathrm{k}_i\nmidsep X_i, \Theta_i\right)\left(x,\theta\right)$ & $q_\theta q_\mathrm{f}$\\
        \hline
    \end{tabular}
    \caption{Example control parameterizations fitting in the HpT-F framework of Definition~\protect\ref{def:hptf}.
        One control DOF corresponds to $n_\mathrm{u}$ decision variables in the tube synthesis problem,
        and $q_\mathrm{f}$ is the number of vertices required to represent the set $X_\mathrm{f}$.\label{tab:paras}}
\end{table}

Next, the HpT-S and HpT-F are combined into a single tube.
Such a tube which will be called an HpT-SF (where ``SF'' stands for ``scenario/fixed-complexity'').

\begin{defn}[HpT-SF]\label{def:hpt-sf}
    Let $N\in\mathbb{N}_{[2,\infty)}$ and $N_0\in\mathbb{N}_{[1,N]}$.
    A heterogeneously parameterized tube $\mathbf{T}\in\mathcal{T}_N(x,\mathbf{\Theta}|\mathcal{P}_N)$ satisfying Definition~\ref{def:hpt} is an HpT-SF,
    if it can be decomposed as $\mathbf{T} = \left(\mathbf{T}^0, \mathbf{T}^1\right)$ with
    \begin{equation}\label{eq:split}
        \begin{aligned}
            \mathbf{T}^0 &= \left(\seq{X^0}{0}{N_0-1}, \seq{K^0}{0}{N_0-1}\right),\\
            \mathbf{T}^1 &= \left(\seq{X^1}{N_0}{N}, \seq{K^1}{N_0}{N-1}\right),
        \end{aligned}
    \end{equation}
    where $\mathbf{T}^0$ is an HpT-S according to Definition~\ref{def:hpts} and $\mathbf{T}^1$ is an HpT-F according to Definition~\ref{def:hptf}.
\end{defn}

\begin{figure*}[t]
    \centering
    \includegraphics[width=0.85\plotwidthfull]{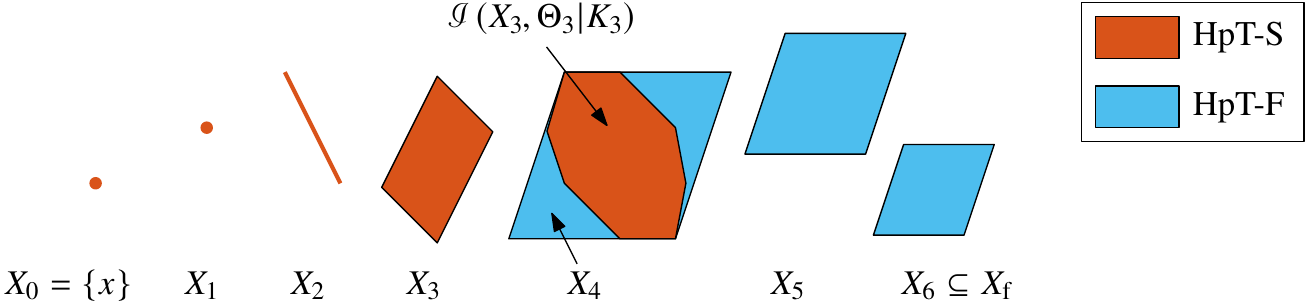}
    \caption{Example of a HpT-SF prediction structure in a two-dimensional state space ($N=6$).\label{fig:hpt-sf}}
    \vspace{1ex}\hrule
\end{figure*}

In the above definition, since both sections $\mathbf{T}^0$ and $\mathbf{T}^1$ are heterogeneously parameterized according to Definition~\ref{def:hpt},
it follows directly that the same holds for the complete tube $\mathbf{T}=\left(\mathbf{T}^0,\mathbf{T}^1\right)$.
To clarify the concept, a graphical representation of an example HpT-SF is given in Figure~\ref{fig:hpt-sf}.
For the first prediction time instances, the HpT-S structure is employed.
The number of vertices of these cross sections doubles at every prediction step after the first one.
Then, at prediction step $N_0=4$, a transition is made to the HpT-F structure: from this point on, the complexity of the tube cross sections remains constant as they are all homothetic to the same polytope.

The next proposition shows that the HpT-SF prediction structure satisfies Assumption~\ref{ass:hptmpc-rfeas}, and therefore leads to a recursively feasible MPC as proven in Theorem~\ref{thm:props}.

\begin{prop}\label{prop:rfeas}
    Suppose Assumption~\ref{ass:implementable} is satisfied.
    Then, the HpT-SF structure of Definition~\ref{def:hpt-sf} satisfies Assumption~\ref{ass:hptmpc-rfeas}.(i).
\end{prop}
\begin{pf}
    First, consider the HpT-S part $\mathbf{T}^0$.
    In reference to Lemma~\ref{lem:succ}, recall that $X^+_0 \subseteq X_1$ is a singleton.
    Applying the first construction of Remark~\ref{rem:succ} for $i\in\mathbb{N}_{[0,N_0-1]}$ (and noting that $\Theta^+_0 = \{\theta(k+1)\}$),
    gives a sequence $\seq{X^+}{0}{N_0-1}$ where the amount of vertices of the $i$-th set equals $q_\mathrm{x}(i)=q_\theta^{\mathrm{max}\{0,i-1\}}$.
    This is in accordance with Definition~\ref{def:hpts}: hence, for all $i\in\mathbb{N}_{[0,N_0-1]}$, there exists a $p^{\mathrm{x}+}_i:\ P^\mathrm{x}\bigl(p^{\mathrm{x}+}_i\nmidsep i\bigr) = X^+_i$.
    Next, for all $i\in\mathbb{N}_{[0,N_0-1]}$, the restricted controllers $K^+_i \in \restr{K_{i+1}}{X^+_i}{\Theta^+_i}$ are the vertex controllers on the sets $X^+_i$, which also agrees with Definition~\ref{def:hpts}.
    Because $X^+_i\subseteq X_{i+1}$, the corresponding vertex control actions in $p^{\mathrm{k}+}_i\in\mathbb{P}(i)$ can be taken as convex combinations of the elements of $p^\mathrm{k}_{i+1}\in\mathbb{P}(i+1)$.
    Therefore, for all $i\in\mathbb{N}_{[0,N_0-1]}$, the existence of $p^{\mathrm{k}+}_i\in\mathbb{P}^\mathrm{k}(i)$ such that $P\bigl(p^{\mathrm{k}+}_i\nmidsep i\bigr) = K^+_i$ is guaranteed.
    Second, consider the HpT-F part $\mathbf{T}^1$.
    Applying the second construction of Remark~\ref{rem:succ} for $i\in\mathbb{N}_{[N_0,N-1]}$ gives the sequence $\seq{X^+}{N_0}{N}$ where $\seq{X^+}{N_0}{N-2} = \seq{X}{N_0+1}{N-1}$,
    $X^+_{N-1} = \gamma X_\mathrm{f}$, and $X^+_N=\lambda\gamma X_\mathrm{f}$.
    Because all sets in $\seq{X^+}{N_0}{N}$ can be represented as $X^+_i = z_i\oplus\alpha_iX_\mathrm{f}$, it follows immediately from Definition~\ref{def:hptf} that
    for all $i\in\mathbb{N}_{[N_0,N]}$, there exists a $p^{\mathrm{x}+}_i:\ P^\mathrm{x}\bigl(p^{\mathrm{x}+}_i\nmidsep i\bigr) = X^+_i$.
    From this construction, it also follows that for $i\in\mathbb{N}_{[N_0,N-2]}$, $K^+_i = K_{i+1} \in \restr{K_{i+1}}{X^+_i}{\Theta^+_i}$
    and that $K^+_{N-1} \in \restr{K_\mathrm{f}}{\gamma X_\mathrm{f}}{\Theta^+_{N-1}}$.
    Thus, Definition~\ref{def:hptf} guarantees that, for all $i\in\mathbb{N}_{[N_0,N-1]}$, there exist $p^{\mathrm{k}+}_i: P\bigl(p^{\mathrm{k}+}_i|i\bigr) = K^+_i$.
    This concludes the proof.
    \qed
\end{pf}

The HpT-F part in the HpT-SF structure can be implemented similarly as in \cite{Hanema2016:cdc:final}.
Hence, for fixed $N_0$, the number of variables and constraints in the tube synthesis \eqref{eq:tube-synth} will be in the order of $O(N)$.
For variable $N_0$, the complexity of the tube synthesis problem is in the order $O(q_{\theta}^{N_0})$:
a remaining question is how to select $N_0$.
To obtain the least conservative control law for a given prediction horizon $N$, one can choose $N_0\leq N$ as large as computational resources allow.
Another approach is to compare the complexity of an HpT-S of length $N_0$ with the complexity of an HpT-F of the same length.
The last cross-section $X_{N_0-1}$ of an HpT-S of length $N_0$ has $q_\theta^{N_0-2}$ vertices.
Then, if the set $X_\mathrm{f}$ from Definition~\ref{def:hptf} has $q_\mathrm{f}$ vertices,
one can select $N_0$ such that $q^{N_0-2}_\theta \approx q_\mathrm{f}$, i.e.,
\begin{equation}\label{eq:n0}
    N_0 = \mathrm{round}\left(\frac{\log q_\mathrm{f}}{\log q_\theta}\right) + 2.
\end{equation}
The value \eqref{eq:n0} is the approximate value where the complexity---in terms of the number of vertices of the cross sections---of the HpT-S part grows beyond that of the HpT-F part,
so that it is advantageous to switch to the HpT-F parameterization after $N_0$ prediction steps.

Note that the tractable parameterizations in this sections were based on tubes with polytopic cross sections.
In the presented framework, it would be possible to develop parameterizations that use ellipsoidal cross sections.
Generally, the representation complexity of ellipsoids scales better in the state dimension,
but synthesizing ellipsoidal tubes requires solving an semi-definite program (SDP) instead of a linear program (LP).


\section{Numerical examples}
\label{sec:numex}

In this section, two numerical examples are provided to demonstrate the HpTMPC algorithm.
It is shown how different choices in the parameterization structure affect properties such as the DOA size and computation time.


\subsection{Parameter-varying double integrator}
\label{sec:numex:ex1}

Consider the LPV-SS representation
\begin{multline*}
    x(k+1) = \Biggl(\begin{bmatrix}1 & 1\\0 & 1\end{bmatrix} + \begin{bmatrix}0.1 & 0\\0 & 0.1\end{bmatrix}\theta_1(k) +
    \begin{bmatrix}0.5 & 0.5\\0 & 0\end{bmatrix}\theta_2(k)\\ + \begin{bmatrix}0 & 0\\0 & 0.2\end{bmatrix}\theta_3(k)\Biggr)x(k) +
    \begin{bmatrix}0.5\\1\end{bmatrix}u(k)
\end{multline*}
with the constraint and scheduling sets given as
\begin{multline*}
    \mathbb{X} = \left\{x\in\mathbb{R}^2\midsep \|x\|_\infty\leq 6\right\},\\
    \mathbb{U} = \left\{u\in\mathbb{R}\midsep |u|\leq 1\right\},
    \Theta = \left\{\theta\in\mathbb{R}^3\midsep \|\theta\|_\infty \leq 1\right\}.
\end{multline*}
The set $\Theta$ is a hypercube in 3 dimensions and therefore has 8 vertices.
The purpose of this example is to demonstrate the effect of changing the heterogeneous parameterization structure.
Therefore, all other design parameters are kept fixed based on the following choices:

\begin{itemize}
    \item The prediction horizon is set to $N=10$.

    \item The scheduling tubes $\mathbf{\Theta}_k\subseteq\Theta^N$ are constructed such that $\mathbf{\Theta}_k=\{\{\theta(k)\},\Theta,\dots,\Theta\}$ for all $k$.
    In this way, attention is focused on the effect of changing the heterogeneous parameterization structure, and not on different possible ways of constructing a scheduling tube
    (see Section~\ref{sec:prelim:anticipative} for some possible constructions).
    
    \item The tuning parameters are set to $Q=I$ and $R=1$.
    
    \item The terminal set $X_\mathrm{f}$ is computed to be $0.95$-contractive with respect to a robust LTI terminal controller $K_\mathrm{f}(x,\theta) = K_\mathrm{f}x$.
    This set is computed using \cite{Miani2005} and is described by $10$ vertices.
    The restriction to an LTI terminal controller in this case yields a set $X_\mathrm{f}$ with a relatively small volume, but also with a relatively low number of vertices.
    Furthermore, in this example, the relatively small volume of the set allows to illustrate more clearly the effect of the tube parameterization on the DOA of the resulting controllers.

    \item The maximal ``worst-case'' DOA is $\mathcal{X}^0_N = \mathcal{X}^\mathrm{max}_N(\mathbf{\Theta}^\mathrm{worst})$ where $\mathbf{\Theta}^\mathrm{worst}=\{\Theta,\dots,\Theta\}$.
    This corresponds to the largest DOA that can be achieved by any possible controller. It is also computed using \cite{Miani2005}.
\end{itemize}

Three tube-based controllers based on different parameterization structures $\mathcal{P}_N$ are compared in terms of the achieved DOA and the number of required control DOFs:

\begin{description}
    \item[Design 1] ($\mathcal{X}_N^1$) Homothetic tube with vertex controls, i.e., a tube satisfying Definition~\ref{def:hptf} with control parameterization 3 from Table~\ref{tab:paras}.
    This is the same as the controller in \cite{Hanema2016:cdc:final}.
    
    \item[Design 2] ($\mathcal{X}_N^2$) Homothetic tube with ``simple'' control parameterization,
    i.e., a tube satisfying Definition~\ref{def:hptf} with parameterization 1 from Table~\ref{tab:paras}.
    
    \item[Design 3] ($\mathcal{X}_N^3$) A heterogeneous design, consisting of a scenario tree for the first $3$ prediction time instances,
    a homothetic tube with vertex control parameterization for the next $3$ instances,
    and a homothetic tube with the simple control parameterization 1 of Table~\ref{tab:paras} for the remaining prediction steps.
\end{description}

\begin{figure}
    \centering
    \includegraphics[width=\plotwidth,trim={0.8cm 0 1.45cm 0.1cm},clip]{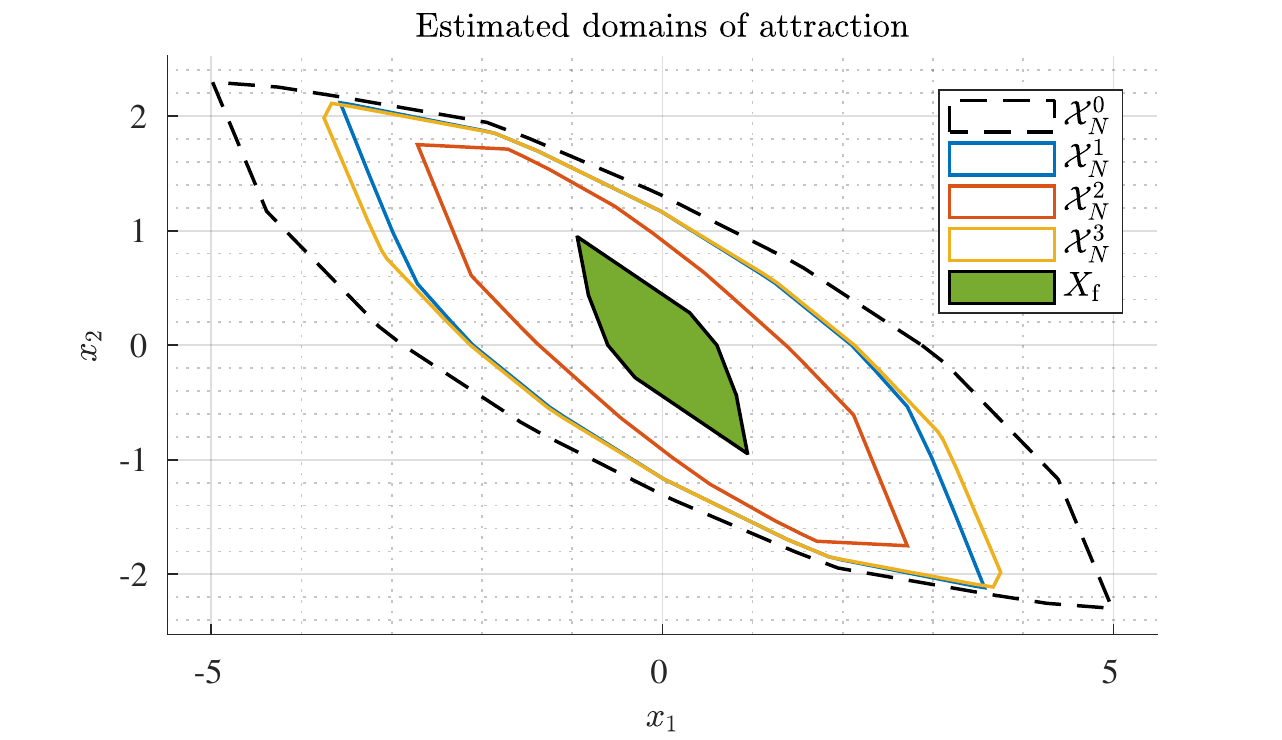}
    \caption{Realized (estimated) DOAs for the three designs in Example~1.
    Here, $\mathcal{X}_N^0 = \mathcal{X}^\mathrm{max}_N(\mathbf{\Theta}^\mathrm{worst})$, whereas $\mathcal{X}_N^i$ is the DOA corresponding to Design~$i$ for $i\in\{1,2,3\}$.\label{fig:ex1-doa}}
\end{figure}

\begin{figure}
    \centering
    \includegraphics[width=\plotwidth,trim={0.3cm 0 1cm 0},clip]{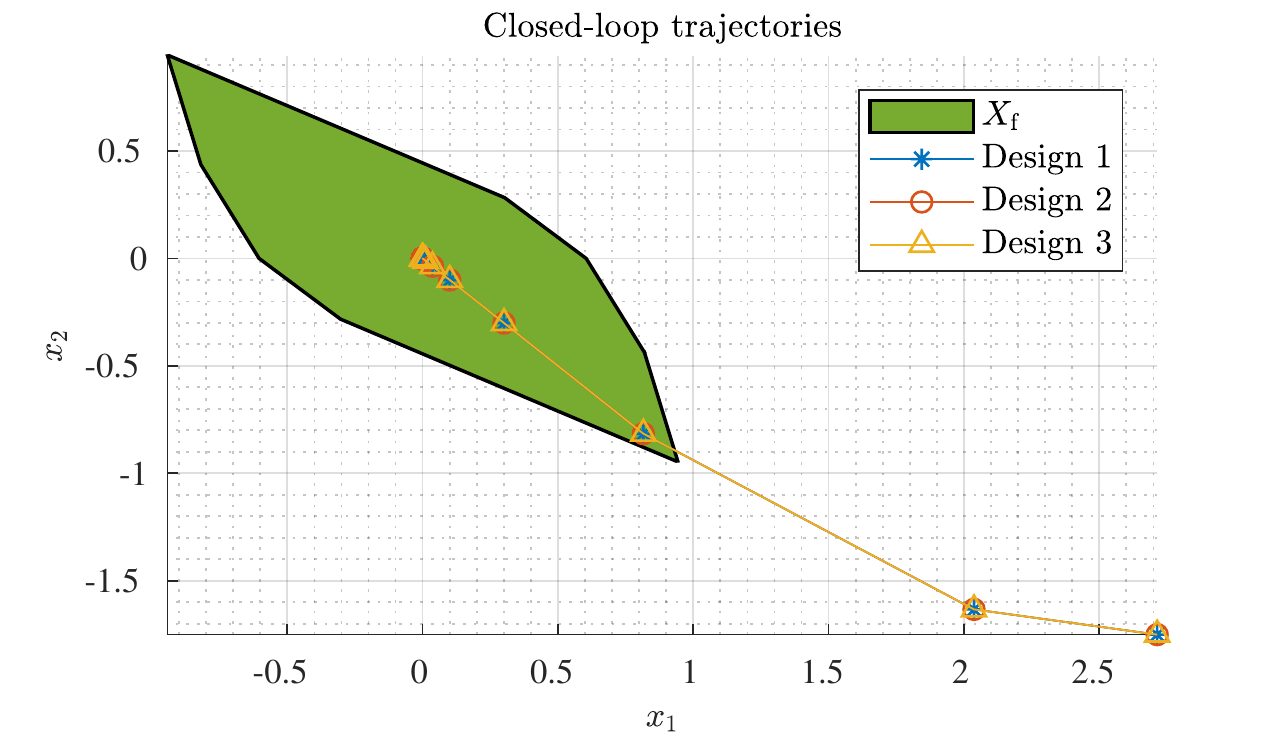}\\
    \includegraphics[width=\plotwidth,trim={0.3cm 0 1cm 0},clip]{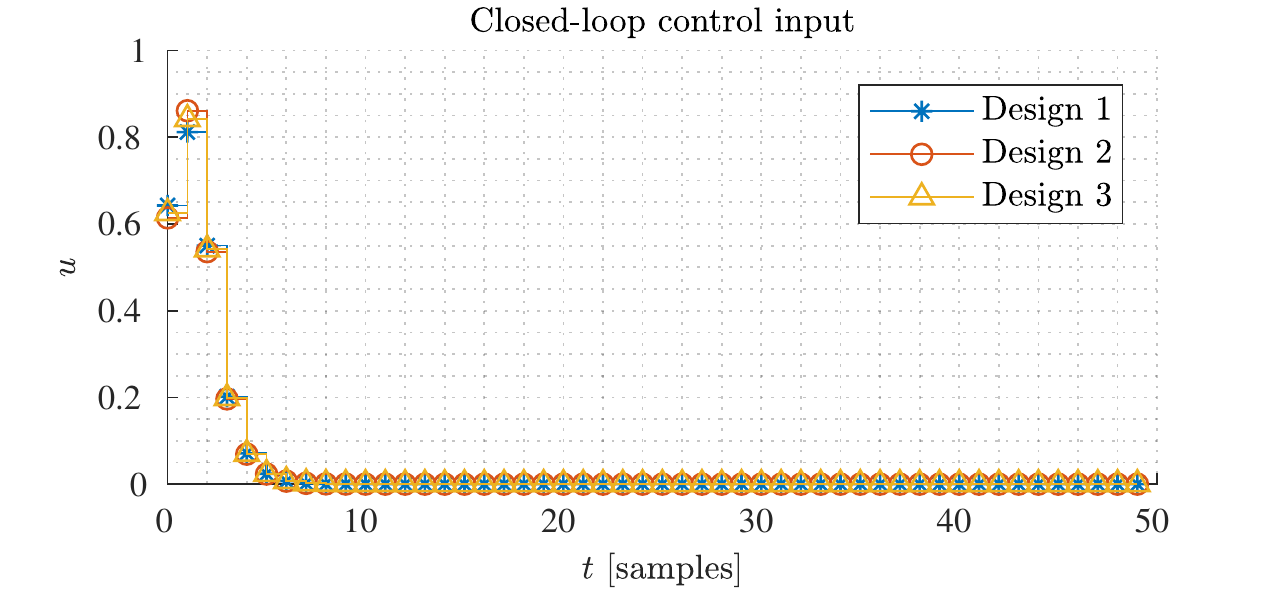}
    \caption{Illustrative closed loop trajectories in Example~1.\label{fig:ex1-cl}}
\end{figure}

The realized DOAs for the three designs were approximated by gridding the state space\footnote{Exact computation of the DOAs using multi-parametric linear programming is intractable given the sizes of the LPs.} and are displayed in Figure~\ref{fig:ex1-doa}.
These DOAs are all computed with respect to a scheduling tube $\mathbf{\Theta}^\mathrm{worst}=\{\Theta,\dots,\Theta\}$.
The associated set volumes and the number of control DOF are displayed in Table~\ref{tab:ex1-comparison}.
Closed-loop trajectories for an initial state at which all the designs are feasible is shown in Figure~\ref{fig:ex1-cl}.
The scheduling trajectory was a randomly time-varying signal generated by drawing, at each instant $k$, a value $\theta(k)$ uniformly from $\Theta$.
For this initial state, the closed-loop inputs are slightly different, but the resulting state trajectories are virtually indistinguishable.
\begin{table}
    \centering
    \begin{tabular}{lcl}
        Design & DOA vol. & DOF\\\hline
        1 ($\mathcal{X}_N^1$) & $12.5$ & $1+(N-1)q_\theta q_\mathrm{f} = 721$\\
        2 ($\mathcal{X}_N^2$) & $7.51$ & $N=10$\\
        3 ($\mathcal{X}_N^3$) & $13.2$ & $\left(1+q_\theta+q_\theta^2\right) + 3q_\theta q_\mathrm{f} + 4 = 317$\\\hline
    \end{tabular}
    \caption{Comparison of the three designs in Example~1. Note $q_\mathrm{f}$ is the number of vertices of the terminal set, to which the tube cross sections in the HpT-F part are homothetic.\label{tab:ex1-comparison}}
\end{table}

For illustration of the resulting computational complexity of the different designs, a summary of the computation times for the closed-loop simulations is given in Table~\ref{tab:ex1-times}.
Note that Design~2 has only $10$ DOF, but that the complexity of the tube synthesis problem is dominated by the number of constraints necessary to verify the tube set inclusions in that case.
The simulations were executed on a computer with a Intel Core i7-4790 processor at 3.60 GHz and 8 GB RAM, with Gurobi 7.0.2 being the LP solver.

\begin{table}
    \centering
	\begin{tabular}{lccc}
    	Design & Avg. time [ms] & Max. time [ms]\\\hline
    	1 ($\mathcal{X}_N^1$) & 55 & 69\\
    	2 ($\mathcal{X}_N^2$) & 42 & 45\\
        3 ($\mathcal{X}_N^3$) & 37 & 44\\\hline
	\end{tabular}
	\caption{Solver time per sample in the simulation of Example~1.\label{tab:ex1-times}}
\end{table}

Out of the three compared designs, the heterogeneously parameterized controller achieves the largest DOA volume while requiring the
lowest average computation time.
Therefore, this example has shown that the proposed heterogeneous tube parameterization has the potential of improving the trade-off between computational complexity and control performance.

It is emphasized that the DOAs in this example were computed with respect to all possible initial scheduling values $\theta\in\Theta$.
For a specific initial scheduling value $\theta(k)$, the tube synthesis might also be feasible for initial states outside of these domains.
Also, the scheduling tubes were constructed in a ``worst-case'' sense:
whenever knowledge is available to construct refined scheduling tubes $\mathbf{\Theta}\subseteq\Theta^N$ (see Section~\ref{sec:prelim:anticipative}), the DOAs can become larger.


\subsection{Parameter-varying third-order system}
\label{sec:numex:ex2}

Consider a third-order LPV system described by
\begin{align*}
    x(k+1)
    &= \left(I + \tau\begin{bmatrix}0 & 1 & 0\\-0.7\theta_1(k) & -0.4 & 0.2\\0 & -0.3 & -0.1\theta_2(k)\end{bmatrix}\right)x(k)\\
    &+ \tau\begin{bmatrix}0 & 0 & 1\end{bmatrix}^\top u(k)
\end{align*}
and with the constraint and scheduling sets
\begin{align*}
    \mathbb{X} &= \left\{x\in\mathbb{R}^3\midsep |x_1|\leq 0.5, |x_2|\leq 0.1, |x_3|\leq 0.2\right\},\\
    \mathbb{U} &= \left\{u\in\mathbb{R}\midsep |u|\leq 0.2\right\},\\
    \Theta &= \left\{\theta\in\mathbb{R}^2\midsep \theta_1\in [0.5, 1.5], \theta_2\in[0.8,1.2] \right\}.
\end{align*}
In what follows, the sampling time is set to $\tau = 0.36$ s.
Similar to the previous example,
all design parameters except the parameterization structure are kept fixed:

\begin{itemize}
    \item The prediction horizon is set to $N=8$.

    \item The scheduling tubes $\mathbf{\Theta}_k\subseteq\Theta^N$ have been constructed such that $\mathbf{\Theta}_k=\{\{\theta(k)\},\Theta,\dots,\Theta\}$ for all $k$.
    As in the previous example, this means that attention is focused on the effect of changing the heterogeneous parameterization structure, and not on different possible scheduling tube
    constructions
    (see Section~\ref{sec:prelim:anticipative}).
    
    \item The tuning parameters are set to $Q=I$ and $R=5$.
    
    \item The terminal set $X_\mathrm{f}$ is computed---using \cite{Miani2005}---to be $0.98$-contractive
    with respect to an LTI terminal controller $K_\mathrm{f}(x,\theta) = K_\mathrm{f}x$.
    This set is described by $48$ vertices and, equivalently, by $28$ hyperplanes.
\end{itemize}

Now,
three tube-based controllers based on different parameterization structures $\mathcal{P}_N$ are compared in terms of the achieved DOA and the number of control DOFs:

\begin{description}
    \item[Design 1] Homothetic tube with vertex controls, i.e., a tube satisfying Definition~\ref{def:hptf} with parameterization 3 from Table~\ref{tab:paras}.
    
    \item[Design 2] Homothetic tube with ``simple'' control parameterization, i.e., a tube satisfying Definition~\ref{def:hptf} with control parameterization 1 from Table~\ref{tab:paras}.
    
    \item[Design 3] A heterogeneous design, consisting of a scenario tree for the first $4$ prediction time instances,
    and a homothetic tube with the simple control parameterization 1 of Table~\ref{tab:paras} for the remaining prediction steps.
\end{description}

\begin{figure}
    \centering
    \includegraphics[width=\plotwidth,trim={0.2cm 0 1.5cm 0.1cm},clip]{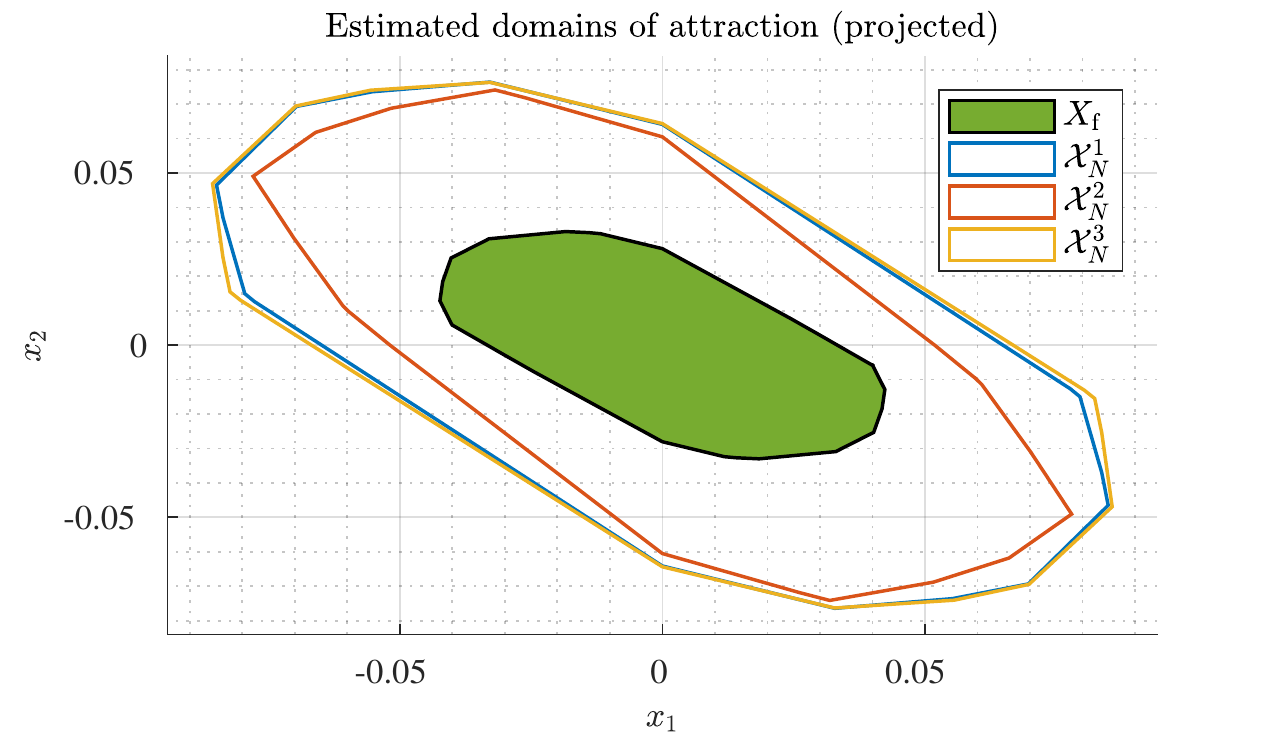}
    \caption{Realized (estimated) DOAs for the three designs in Example~2: projection on the $(x_1,x_2)$-space.\label{fig:ex2-doa}}
\end{figure}
\begin{table}
    \centering
    \begin{tabular}{lcl}
        Design & DOA vol. & DOF\\\hline
        1 ($\mathcal{X}_N^1$) & $3.13\cdot 10^{-3}$ & $1+(N-1)q_\theta q_\mathrm{f} = 1345$\\
        2 ($\mathcal{X}_N^2$) & $2.43\cdot 10^{-3}$ & $N=8$\\
        3 ($\mathcal{X}_N^3$) & $3.23\cdot 10^{-3}$ & $\left(1+q_\theta+q_\theta^2+q_\theta^3\right) + 4 = 95$\\\hline
    \end{tabular}
    \caption{Comparison of the three designs in Example~2.\label{tab:ex2-comparison}}
\end{table}

A projection on the $(x_1,x_2)$-space of the estimated DOAs realized by the three designs is shown in Figure~\ref{fig:ex2-doa}
(plotting the full three-dimensional sets would be illegible).
The volumes of the estimated DOAs and the number of control DOFs are summarized in Table~\ref{tab:ex2-comparison}.
An illustration of the computation times required to solve the tube synthesis problems for the three different designs is given in Table~\ref{tab:ex2-times}.
These times were obtained by simulating the closed-loop system with an initial state $x_0=\left[0.05\ \ 0\ \ 0\right]^\top$.
As remarked before, it should be possible to improve these times by considering a more efficient linear programming implementation.

\begin{table}
    \centering
	\begin{tabular}{lccc}
    	Design & Avg. time [ms] & Max. time [ms]\\\hline
    	1 ($\mathcal{X}_N^1$) & 613 & 778\\
    	2 ($\mathcal{X}_N^2$) & 310 & 342\\
        3 ($\mathcal{X}_N^3$) & 317 & 360\\\hline
	\end{tabular}
	\caption{Solver time per sample in the simulation of Example~2.\label{tab:ex2-times}}
\end{table}

Out of the compared designs, the heterogeneously parameterized controller achieves the largest DOA volume, while
requiring a computation time that is only slightly higher than the ``simple'' controller.
As in the previous example, this confirms that the heterogeneous tube parameterization proposed in this paper provides an
improved trade-off between computational complexity and control performance as measured by the DOA volume.


\section{Concluding remarks}
\label{sec:conclusion}

In this paper, a framework for the construction of MPC schemes for LPV-SS models was developed,
based on the construction of so-called heterogeneously parameterized tubes.
Possibilities for future research include the extension of the framework to handle LPV models also affected by additive disturbances,
the implementation of tube parameterizations based on ellipsoids,
and the investigation of algorithmic approaches to systematically design heterogeneous parameterization structures.


\bibliographystyle{IEEEtranLocal}
\bibliography{library_thesis_local}

\begin{thebibliography}{10}
\def\url#1{}
\csname url@samestyle\endcsname
\providecommand{\newblock}{\relax}
\providecommand{\bibinfo}[2]{#2}
\providecommand{\BIBentrySTDinterwordspacing}{\spaceskip=0pt\relax}
\providecommand{\BIBentryALTinterwordstretchfactor}{4}
\providecommand{\BIBentryALTinterwordspacing}{\spaceskip=\fontdimen2\font plus
\BIBentryALTinterwordstretchfactor\fontdimen3\font minus
  \fontdimen4\font\relax}
\providecommand{\BIBforeignlanguage}[2]{{%
\expandafter\ifx\csname l@#1\endcsname\relax
\typeout{** WARNING: IEEEtran.bst: No hyphenation pattern has been}%
\typeout{** loaded for the language `#1'. Using the pattern for}%
\typeout{** the default language instead.}%
\else
\language=\csname l@#1\endcsname
\fi
#2}}
\providecommand{\BIBdecl}{\relax}
\BIBdecl

\bibitem{Lee1997}
J.~H. Lee and Z.~Yu, ``{{Worst-case Formulations of Model Predictive Control
  for Systems with Bounded Parameters}},'' \emph{Automatica}, vol.~33, pp.
  763--781, 1997.

\bibitem{RawlingsMayne2009}
J.~B. Rawlings and D.~Q. Mayne, \emph{{{Model Predictive Control: Theory and
  Design}}}.\hskip 1em plus 0.5em minus 0.4em\relax Nob Hill Publishing, 2009.

\bibitem{Rakovic2015a}
S.~V. Rakovi{\'{c}}, ``{{Robust Model-Predictive Control}},'' in
  \emph{{Encyclopedia of Systems and Control}}.\hskip 1em plus 0.5em minus
  0.4em\relax Springer, 2015, pp. 1225--1233.

\bibitem{Lu2000}
Y.~Lu and Y.~Arkun, ``{{Quasi-Min-Max MPC algorithms for LPV systems}},''
  \emph{Automatica}, vol.~36, pp. 527--540, 2000.

\bibitem{Casavola2002}
A.~Casavola, D.~Famularo, and G.~Franz{\`{e}}, ``{{A Feedback Min-Max MPC
  Algorithm for LPV Systems Subject to Bounded Rates of Change of
  Parameters}},'' \emph{IEEE Transactions on Automatic Control}, vol.~47, pp.
  1147--1153, 2002.

\bibitem{Casavola2007}
A.~Casavola, D.~Famularo, G.~Franz{\`{e}}, and E.~Garone, ``{{A dilated MPC
  control strategy for LPV linear systems}},'' in \emph{{Proc. of the 2007
  European Control Conference}}, 2007, pp. 460--466.

\bibitem{Casavola2008a}
A.~Casavola, D.~Famularo, and G.~Franz{\`{e}}, ``{{A predictive control
  strategy for norm-bounded LPV discrete-time systems with bounded rates of
  parameter change}},'' \emph{Int. J. of Robust and Nonlinear Control},
  vol.~18, pp. 714--740, 2008.

\bibitem{Zheng2013}
P.~Zheng, D.~Li, Y.~Xi, and J.~Zhang, ``{{Improved model prediction and RMPC
  design for LPV systems with bounded parameter changes}},'' \emph{Automatica},
  vol.~49, pp. 3695--3699, 2013.

\bibitem{Bacic2003}
M.~Bacic, M.~Cannon, Y.~I. Lee, and B.~Kouvaritakis, ``{{General Interpolation
  in MPC and Its Advantages}},'' \emph{IEEE Transactions on Automatic Control},
  vol.~48, pp. 1092--1096, 2003.

\bibitem{Pluymers2005f}
B.~Pluymers, J.~A.~K. Suykens, and B.~{De Moor}, ``{{Min-max feedback MPC using
  a time-varying terminal constraint set and comments on ``Efficient robust
  constrained model predictive control with a time-varying terminal constraint
  set''}},'' \emph{Systems {\&} Control Letters}, vol.~54, pp. 1143--1148,
  2005.

\bibitem{Wan2006}
Z.~Wan, B.~Pluymers, M.~V. Kothare, and B.~{De Moor}, ``{{Comments on:
  ``Efficient robust constrained model predictive control with a time varying
  terminal constraint set'' by Wan and Kothare}},'' \emph{Automatica}, vol.~55,
  pp. 618--621, 2006.

\bibitem{Kouvaritakis2000}
B.~Kouvaritakis, J.~A. Rossiter, and J.~Schuurmans, ``{{Efficient robust
  predictive control}},'' \emph{IEEE Transactions on Automatic Control},
  vol.~45, pp. 1545--1549, 2000.

\bibitem{Cannon2005}
M.~Cannon and B.~Kouvaritakis, ``{{Optimizing prediction dynamics for robust
  MPC}},'' \emph{IEEE Transactions on Automatic Control}, vol.~50, pp.
  1892--1897, 2005.

\bibitem{Hanema2016:cdc:final}
J.~Hanema, R.~T{\'{o}}th, and M.~Lazar, ``{{Tube-based anticipative model
  predictive control for linear parameter-varying systems}},'' in \emph{{Proc.
  of the 55th IEEE Conference on Decision and Control}}, 2016, pp. 1458--1463.

\bibitem{Langson2004}
W.~Langson, I.~Chryssochoos, S.~V. Rakovi{\'{c}}, and D.~Q. Mayne, ``{{Robust
  model predictive control using tubes}},'' \emph{Automatica}, vol.~40, pp.
  125--133, 2004.

\bibitem{Mayne2005}
D.~Q. Mayne, M.~M. Seron, and S.~V. Rakovi{\'{c}}, ``{{Robust model predictive
  control of constrained linear systems with bounded disturbances}},''
  \emph{Automatica}, vol.~41, pp. 219--224, 2005.

\bibitem{Rakovic2012a}
S.~V. Rakovi{\'{c}}, B.~Kouvaritakis, R.~Findeisen, and M.~Cannon,
  ``{{Homothetic tube model predictive control}},'' \emph{Automatica}, vol.~48,
  pp. 1631--1638, 2012.

\bibitem{Rakovic2012}
S.~V. Rakovi{\'{c}}, B.~Kouvaritakis, M.~Cannon, C.~Panos, and R.~Findeisen,
  ``{{Parameterized tube model predictive control}},'' \emph{IEEE Transactions
  on Automatic Control}, vol.~57, pp. 2746--2761, 2012.

\bibitem{Rakovic2016a}
S.~V. Rakovi{\'{c}}, W.~S. Levine, and B.~A{\c{c}}ıkmeşe, ``{{Elastic Tube
  Model Predictive Control}},'' in \emph{{Proc. of the 2016 American Control
  Conference}}, 2016, pp. 3594--3599.

\bibitem{Munoz-Carpintero2013a}
D.~Mu{\~{n}}oz-Carpintero, M.~Cannon, and B.~Kouvaritakis, ``{{Robust MPC
  strategy with optimized polytopic dynamics for linear systems with additive
  and multiplicative uncertainty}},'' \emph{Systems {\&} Control Letters},
  vol.~81, pp. 34--41, 2015.

\bibitem{Fleming2015}
J.~Fleming, B.~Kouvaritakis, and M.~Cannon, ``{{Robust Tube MPC for Linear
  Systems With Multiplicative Uncertainty}},'' \emph{IEEE Transactions on
  Automatic Control}, vol.~60, pp. 1087--1092, 2015.

\bibitem{Brunner2013}
F.~D. Brunner, M.~Lazar, and F.~Allg{\"{o}}wer, ``{{An Explicit Solution to
  Constrained Stabilization via Polytopic Tubes}},'' in \emph{{Proc. of the
  52nd IEEE Conference on Decision and Control}}, 2013, pp. 7721--7727.

\bibitem{Hanema2017:aut:final}
J.~Hanema, M.~Lazar, and R.~T{\'{o}}th, ``{{Stabilizing Tube-Based Model
  Predictive Control: Terminal Set and Cost Construction for LPV Systems}},''
  \emph{Automatica}, vol.~85, pp. 137--144, 2017.

\bibitem{Goulart2008}
P.~J. Goulart, E.~C. Kerrigan, and D.~Ralph, ``{{Efficient robust optimization
  for robust control with constraints}},'' \emph{Mathematical Programming},
  vol. 114, pp. 115--147, 2008.

\bibitem{Lucia2013}
S.~Lucia, T.~Finkler, and S.~Engell, ``{{Multi-stage nonlinear model predictive
  control applied to a semi-batch polymerization reactor under uncertainty}},''
  \emph{Journal of Process Control}, vol.~23, pp. 1306--1319, 2013.

\bibitem{Maiworm2015}
M.~Maiworm, T.~B{\"{a}}thge, and R.~Findeisen, ``{{Scenario-based Model
  Predictive Control: Recursive Feasibility and Stability}},'' in \emph{{Proc.
  of the 9th IFAC Symposium on Advanced Control of Chemical Processes}}, 2015,
  pp. 50--56.

\bibitem{Scokaert1998}
P.~O.~M. Scokaert and D.~Q. Mayne, ``{{Min-max Feedback Model Predictive
  Control for Constrained Linear Systems}},'' \emph{IEEE Transactions on
  Automatic Control}, vol.~43, pp. 1136--1142, 1998.

\bibitem{Kerrigan2004}
E.~C. Kerrigan and J.~M. Maciejowski, ``{{Feedback min-max model predictive
  control using a single linear program: robust stability and the explicit
  solution}},'' \emph{Int. J. of Robust and Nonlinear Control}, vol.~14, pp.
  395--413, 2004.

\bibitem{MunozDeLaPena2005}
D.~{Mu{\~{n}}oz de la Pe{\~{n}}a}, T.~Alamo, D.~R. Ramirez, and E.~F. Camacho,
  ``{{Min-max model predictive control as a quadratic program}},'' in
  \emph{{Proc. of the 16th IFAC World Congress}}, 2005, pp. 263--268.

\bibitem{MunozDeLaPena2006}
D.~{Mu{\~{n}}oz de la Pe{\~{n}}a}, T.~Alamo, A.~Bemporad, and E.~F. Camacho,
  ``{{Feedback Min-Max Model Predictive Control Based on a Quadratic Cost
  Function}},'' in \emph{{Proc. of the 2006 American Control Conference}},
  2006, pp. 1575--1580.

\bibitem{Subramanian2018a}
S.~Subramanian, S.~Lucia, S.~A.~B. Birjandi, R.~Paulen, and S.~Engell, ``{{A
  Combined Multi-stage and Tube-based MPC Scheme for Constrained Linear
  Systems}},'' in \emph{{Proc. of the 6th IFAC Conference on Nonlinear Model
  Predictive Control}}, 2018, pp. 481--486.

\bibitem{Aeyels1998}
D.~Aeyels and J.~Peuteman, ``{{A New Asymptotic Stability Criterion for
  Nonlinear Time-Variant Differential Equations}},'' \emph{IEEE Transactions on
  Automatic Control}, vol.~43, pp. 968--971, 1998.

\bibitem{Jiang2002}
Z.~P. Jiang and Y.~Wang, ``{{A converse Lyapunov theorem for discrete-time
  systems with disturbances}},'' \emph{Systems {\&} Control Letters}, vol.~45,
  pp. 49--58, 2002.

\bibitem{Hanema2017:cdc:final}
J.~Hanema, R.~T{\'{o}}th, and M.~Lazar, ``{{Stabilizing Non-linear MPC using
  Linear Parameter-Varying Representations}},'' in \emph{{Proc. of the 56th
  IEEE Conference on Decision and Control}}, 2017, pp. 3582--3587.

\bibitem{Abelson1996}
H.~Abelson, G.~J. Sussman, and J.~Sussman, \emph{{{Structure and Interpretation
  of Computer Programs}}}, 2nd~ed.\hskip 1em plus 0.5em minus 0.4em\relax MIT
  Press, 1996.

\bibitem{Mayne2000}
D.~Q. Mayne, J.~B. Rawlings, C.~V. Rao, and P.~O.~M. Scokaert, ``{{Constrained
  model predictive control: Stability and optimality}},'' \emph{Automatica},
  vol.~36, pp. 789--814, 2000.

\bibitem{Blanchini2007}
F.~Blanchini, S.~Miani, and C.~Savorgnan, ``{{Stability results for linear
  parameter varying and switching systems}},'' \emph{Automatica}, vol.~43, pp.
  1817--1823, 2007.

\bibitem{Lucia2014}
S.~Lucia, R.~Paulen, and S.~Engell, ``{{Multi-stage Nonlinear Model Predictive
  Control with verified robust constraint satisfaction}},'' in \emph{{Proc. of
  the IEEE Conference on Decision and Control}}, 2014, pp. 2816--2821.

\bibitem{Miani2005}
S.~Miani and C.~Savorgnan, ``{{MAXIS-G: A software package for computing
  polyhedral invariant sets for constrained LPV systems}},'' in \emph{{Proc. of
  the 44th IEEE Conference on Decision and Control, and the European Control
  Conference}}, 2005, pp. 7609--7614.

\end{thebibliography}


\end{document}